\documentclass[a4paper,11pt]{article}

\usepackage{jheppub2} 
\usepackage[T1]{fontenc} % if needed
\usepackage{enumitem}
\usepackage{bbm}
\usepackage{xcolor}
\usepackage[utf8]{inputenc}
\usepackage{cleveref}
\usepackage{amsmath}
\usepackage{amssymb}
\usepackage{pifont}
\usepackage{bm}
\usepackage{enumitem}
\usepackage{makecell}
\usepackage{subfig} 
\usepackage{tabularx,booktabs}
\newcolumntype{C}{>{\centering\arraybackslash}X}
\usepackage{graphicx}
\usepackage{multirow}
\usepackage[normalem]{ulem}
\usepackage{comment}
\usepackage{booktabs}
\usepackage{slashed}
\usepackage[table, svgnames, dvipsnames]{xcolor}
\usepackage{hyphenat} 
\usepackage{multirow,bigdelim}
\usepackage{bbold}

\usepackage{lipsum}
\usepackage{makecell}
\usepackage{mathtools,slashed}
\usepackage{cancel}
\usepackage{blkarray}

\newcommand{\thetanu}{\theta^{\nu}}
\newcommand{\thetal}{\theta^{l}}
\newcommand{\Ul}{U_l}

\newcommand{\UPMNS}{U_{\rm PMNS}}

\usepackage{tikz-feynman}
\tikzfeynmanset{compat=1.1.0}
\setlist[enumerate]{wide=0pt, widest=99,leftmargin=\parindent, labelsep=*}

\newcommand{\med}[1]{\langle #1\rangle}

\newcommand{\Uf}[1]{\text{U(#1)}_\mathrm{F}}
\newcommand{\SUf}[1]{\text{SU(#1)}_\mathrm{F}}

\newdimen\arrowsize
\newdimen\mylw
\pgfkeys{/my arrows/chemeq/.style={draw,thick,double distance=2pt,onearc-onearc}}
\pgfkeys{/my arrows/size/.code={\pgfsetarrowoptions{onearc}{#1}}}
\def\myalw{.4pt}
\pgfarrowsdeclare{onearc}{onearc}{%
  \mylw=\myalw
  \pgfarrowsleftextend{-\pgfgetarrowoptions{onearc}-.5\mylw}
  \pgfarrowsrightextend{0pt}
}{%
  \pgfsetdash{}{0pt}
  \mylw=\pgflinewidth
  \pgfsetlinewidth{\myalw}
  \advance\arrowsize by.5\pgflinewidth
  \pgfpathmoveto{\pgfpoint{-\pgfgetarrowoptions{onearc}}{-\pgfgetarrowoptions{onearc}-.5\mylw}}%
  \pgfpatharc{180}{90}{\pgfgetarrowoptions{onearc}}
  \pgfusepathqstroke
}
\definecolor{brickred}{cmyk}{0, 0.9, 0.98, 0.45}
\definecolor{emerald}{cmyk}{0.91, 0.25, 0.9, 0.2}

\title{
\Huge \boldmath \color{brickred}
     Dark Matter Stability from Residual Flavour Symmetries
      }
\author[1]{Giacomo Landini,}
\author[2]{Simone Marciano}
\author[1]{and Federico Mescia}
\affiliation[1]{Istituto Nazionale di Fisica Nucleare, Laboratori Nazionali di Frascati, C.P.~13, 00044 Frascati, Italy}
\affiliation[2]{Departament de Física Teòrica, Universitat de València, 46100 Burjassot, Spain, and\\
Instituto de F\'isica Corpuscular (CSIC-Universidad de Valencia),\\
Parc Cientif\'ic UV, C/Catedr\'atico Jos\'e Beltr\'an 2, E-46980 Paterna, Spain}
\emailAdd{giacomo.landini@lnf.infn.it}
\emailAdd{simone.marciano@ific.uv.es}
\emailAdd{federico.mescia@lnf.infn.it}

\abstract{
We propose a model-independent framework in which Dark Matter stability emerges naturally from residual discrete $\mathbb{Z}_N$ symmetries left unbroken by the spontaneous breaking of continuous flavour symmetries, both Abelian and non-Abelian. No \textit{ad hoc} stabilizing symmetry is required, as Dark Matter stability is an inherent consequence of the underlying flavour dynamics. As an explicit example, we construct a prototype model based on a $\SUf{2} \otimes \Uf{1}$ Froggatt--Nielsen symmetry, which successfully reproduces the observed charged lepton mass spectrum and the PMNS mixing matrix, while identifying the lightest sterile neutrino as a stable Dark Matter candidate. Furthermore, we study the Dark Matter phenomenology within our model in the freeze-in regime, including  production channels mediated  by higher-dimensional operators suppressed by the cutoff scale $\Lambda$. Our {analysis shows}  that continuous flavour symmetries can provide a unified, self-consistent framework addressing both the flavour puzzle and the origin, stability, and production of Dark Matter. 
}

\makeatletter
\gdef\@fpheader{}
\makeatother

\begin{document} 
\maketitle
\flushbottom
\newpage

%%%%%%%%%%%%%%%%%%%%%%%%%%%
\section{Introduction}\label{sec:intro}
%%%%%%%%%%%%%%%%%%%%%%%%%%%
{
Despite the remarkable success of the Standard Model (SM) of particle physics, several fundamental questions remain unanswered. One of the most compelling is the nature of Dark Matter (DM)~\cite{Bertone:2016nfn,Bertone:2004pz,Peter:2012rz,Cirelli:2024ssz}, whose existence is firmly established through astrophysical and cosmological observations, yet for which the SM provides no viable candidate. Any extension of the SM aiming to account for DM must not only introduce new degrees of freedom, but also explain why the DM particle is stable on cosmological timescales. In many proposed scenarios, stability is guaranteed by imposing discrete symmetries by hand (see~\cite{Cirelli:2024ssz} and references therein). While phenomenologically successful, such constructions offer little insight into the fundamental origin of DM stability.

A second major open problem concerns the flavour structure of the SM (see, e.g., Ref.~\cite{Feruglio:2015jfa} for a review).  
Although gauge invariance successfully determines the interactions among the SM particles, it provides no understanding of generational hierarchies, flavour mixing, or CP violation. In addition, the observation of non-zero neutrino masses provides compelling evidence for physics beyond the SM, further suggesting that a deeper theory of flavour remains to be uncovered. 

A particularly attractive framework to address the SM flavour puzzle is provided by continuous Abelian or non-Abelian flavour symmetries in the Froggatt--Nielsen (FN) mechanism~\cite{Leurer:1992wg,Leurer:1993gy,Froggatt:1978nt}. In this paradigm, heavy scalar fields (\textit{flavons}), charged under a flavour symmetry but singlets under the SM gauge group, acquire non-zero vacuum expectation values (VEVs). The spontaneous breaking of the flavour symmetry generates hierarchical Yukawa structures through powers of a small expansion parameter, naturally reproducing the observed fermion mass hierarchies and mixing patterns. 
Interestingly, FN constructions are particularly appealing and timely in light of the next generation of neutrino oscillation experiments, including the Deep Underground Neutrino Experiment (DUNE)~\cite{DUNE:2020ypp,DUNE:2020jqi} and Tokai-to-Hyper-Kamiokande (T2HK)~\cite{Hyper-Kamiokande:2018ofw}. These facilities will provide unprecedented sensitivity to the neutrino mass ordering, leptonic CP violation, and the precise values of the mixing angles, $\theta_{ij}$.\footnote{In particular, DUNE is expected to measure $\sin^2\theta_{23}$ with sub-percent precision and achieve a $5\sigma$ sensitivity to leptonic CP violation over a large fraction of the parameter space (see Appendix \ref{app:PMNS} for notations of the mixing matrix $\UPMNS$).} Such measurements will offer powerful and highly non-trivial tests of flavour-symmetry models~\cite{Harrison:1999cf,Harrison:2002er,Xing:2002sw,Vissani:1997pa,Altarelli:2005yx,King:2011ab,Giarnetti:2025idu,Barger:2001yr,Giarnetti:2024y2407,ESSnuSB:2025y2504}. 

In this work,  we take a step further relative to existing FN literature. We explore the possibility that flavour symmetries play a dual role. Besides explaining the origin of fermion masses and mixing parameters, we show that the spontaneous breaking of flavour symmetries can leave residual discrete symmetries that naturally stabilize a DM candidate. In this way, DM stability is not imposed through an \textit{ad hoc} assumption\footnote{See~\cite{Buttazzo:2019mvl} for a detailed study of accidental symmetries in scalar theories, though not related to flavour.}, as for example in models where neutrino masses arise purely at one loop~\cite{Arbelaez:2022ejo,Restrepo:2013aga,Cai:2017jrq,Giarnetti:2023osf,Ma:2006km,Hagedorn:2018spx,Escribano:2020iqq,Beniwal:2020hjc}. In contrast, it emerges as a direct consequence of the same flavour dynamics responsible for the observed structure of the fermion sector\footnote{A complementary approach to stabilizing DM within flavour symmetries is provided by the Minimal Flavour Violation framework~\cite{Batell:2011tc,Lopez-Honorez:2013wla,Mescia:2024rki,Mescia:2026xju}.}. This provides a unified framework in which two of the most pressing shortcomings of the SM, namely the flavour puzzle and the existence of DM, originate from a common theoretical structure. Specifically, we formulate a general framework to host a stable DM particle within an FN flavour model. 
In our setup, the Lagrangian is invariant under a flavour group, and the DM state is protected by a residual subgroup that survives spontaneous symmetry breaking. Crucially, this stabilizing subgroup is not imposed externally, but arises naturally from the underlying flavour dynamics.\footnote{For comparison, Refs.~\cite{Mandal:2023jnv,Calibbi:2015sfa, Alvarado:2017bax, Cheek:2022yof,Babu:2023zni} provide interesting examples in which a DM candidate is embedded in a Froggatt--Nielsen framework, but its stability is enforced through an additional symmetry, imposed by hand.} It is a well-known result that breaking an Abelian $\Uf{1}$ symmetry through the VEV of a scalar carrying an integer charge $Q$ leaves an unbroken discrete $\mathbb{Z}_{|Q|}$ symmetry. For instance, {in Ref.~\cite{AristizabalSierra:2014irc}, a similar approach has been pursued: the origin of a residual $\mathbb{Z}_N$ symmetry from a $\Uf{1}$ flavour group has been investigated, although without delving into details of the flavour structure of the SM fermions and the connections to the low energy neutrino observables.} Here we extend this mechanism to non-Abelian flavour groups and we focus on the flavour structure. In what follows, we focus on the minimal non-Abelian realization, $\SUf{2} \otimes \Uf{1}$, which provides a transparent benchmark that smoothly recovers the Abelian $\Uf{1}$ limit while offering a straightforward path toward more general $\Uf{N}$ constructions.} 

This paper is organized as follows. In Section~\ref{FN-DM}, we introduce the Froggatt--Nielsen framework and the associated flavour symmetries. We then discuss the residual discrete symmetry that emerges after flavour symmetry breaking and stabilizes the DM candidate.   

In Section~\ref{sec:prototype}, we present a prototype realization of the framework. We first discuss the charged lepton sector in Section~\ref{sec:chargedleptons}, followed by the neutrino sector in Section~\ref{sec:neutrinos}. We then perform a numerical analysis and summarize the resulting predictions in Section~\ref{sec:numerical}.
The DM phenomenology is explored in Section~\ref{sec:DMphenomenology}. Finally, our conclusions are presented in Section~\ref{sec:conclusions}. Additional technical details are collected in the appendices. Appendix~\ref{app:PMNS} summarizes our notation and conventions for the PMNS mixing matrix, and Appendix~\ref{app:map} the relation among the effective active neutrino mass matrix and the parameters in the Lagrangian.
%%%%%%%%%%%%%%%%%%%%%%%%%%%
\section{Stable Dark Matter from residual flavour symmetries}
\label{FN-DM}
A particularly attractive framework to address the flavour puzzle is provided by approximate flavour (or horizontal) symmetries. In these constructions, SM fermions carry non-trivial flavour charges, such that  Yukawa interactions are forbidden in the exact symmetry limit. The flavour symmetry is spontaneously broken by the VEVs of scalar fields, known as \textit{flavons}, which act as spurions controlling the size of the effective Yukawa couplings. Within an effective field theory description, Yukawa operators arise from higher-dimensional interactions involving suitable powers of flavon fields and suppressed by a large ultraviolet scale $\Lambda$. Assuming the flavon VEVs to be moderately below $\Lambda$, the observed hierarchies of fermion masses and mixing angles can be naturally generated through powers of the small expansion parameters $\epsilon \equiv \langle \phi \rangle/\Lambda$ (see Eqs.~\eqref{eq:FNlep} and \eqref{eq:FNneut} in the next section for the charged lepton and neutrino Yukawas, respectively). This expansion parameter is generally chosen to be of the order of the Cabibbo angle ($\epsilon \sim \theta_C \approx 0.22$), as this provides in many models precisely the required shift to reproduce both the correct solar angle $\theta_{12}$ and the reactor angle $\theta_{13}$~\cite{Marzocca2013b}. In this way, the flavour structure of the SM is interpreted as a consequence of the spontaneous breaking of an underlying flavour symmetry rather than as a collection of arbitrary Yukawa couplings.
While a wide variety of such models have been proposed (see, e.g., Ref.~\cite{Feruglio:2015jfa} and references therein), a particularly simple and attractive class is based on a $\SUf{2}\otimes\Uf{1}$ flavour symmetry~\cite{Barbieri:1995uv,Barbieri:1997tu,Barbieri:2011ci,Isidori:2012ts}. In particular, a realistic flavour model based on a $\SUf{2}\otimes\Uf{1}$ symmetry was shown to successfully reproduce both the quark and lepton sectors~\cite{Linster:2018avp}, which we adopt in next section  as a starting point to incorporate a viable DM candidate.

Now, before specifying  a concrete model, in this section we highlight the key features of our mechanism, where residual flavour symmetries emerge naturally from a spontaneously broken flavour symmetry within the FN framework. For this purpose, it is sufficient to consider a global or local $\SUf{2} \otimes \Uf{1}$ symmetry equipped with two flavons:  a doublet $\phi_a$ and a singlet $\chi$ with integer charges $Q_\phi$ and $Q_\chi$ 
and the following VEV alignment,\footnote{Other choices of the VEV alignments provide the same result, thanks to an arbitrary  change of basis.}
\begin{equation}
    \med{\chi}=v_\chi\,\qquad \med{\phi_a}=\begin{pmatrix}
        v_\phi \\
        0
    \end{pmatrix}\,.
\end{equation}
A generic field $\psi$, filling a representation $R$ of $\SUf{2}$, with charge $Q$ under $\Uf{1}$, transforms as

\begin{equation}\label{eq:generalU2}
    \psi\to \exp[i(\beta Q+\alpha^aT_R^a)]\,\psi\,,
\end{equation}
where $T_R^a$ are the $\SUf{2}$ generators in the $R$ representation. The transformation in Eq.~\eqref{eq:generalU2} leaves the VEV of the singlet flavon invariant if 

\begin{equation}\label{eq:beta}
    \beta=2\pi n/Q_\chi,\qquad n\in\mathbb{Z}\,.
\end{equation}
On the other hand, the VEV of the doublet is invariant  if the following additional constraints

\begin{equation}\label{eq:alphas}
   \alpha_{1,2}=0\,,\qquad  \frac
   {\alpha_3}{2}=2\pi\left(m -n\frac{Q_\phi}{Q_\chi}\right),\qquad m\in\mathbb{Z}\,,
\end{equation}
hold. 

The two conditions in~\eqref{eq:beta} and~\eqref{eq:alphas}
define a residual $\mathbb{Z}_{|Q_\chi|}$ symmetry, that can be used to stabilize a DM candidate. {Notice that the residual symmetry acts trivially if all the fields present in the model (including $\phi$) have a charge which is an integer multiple of $Q_\chi$, since, in such a case, one can rescale the singlet charge to unity.
Furthermore, while the value of $Q_\phi$ does not change the nature of the residual discrete symmetry, it contributes in determining the corresponding transformation rules of the fields, as discussed below.}
Depending on the choice of $Q_\chi$ one gets different residual symmetries:

\begin{itemize}
    \item $|Q_\chi|=1\,$: no residual symmetry is present. This is the typical scenario for most FN models. The DM cannot be stable unless extra ad-hoc symmetries, unrelated to the flavour structure, are introduced by hand.
    \item $|Q_\chi|=2\,$: the residual symmetry is $\mathbb{Z}_2$. The lightest state odd under $\mathbb{Z}_2$ is stable and can be a potentially viable DM candidate. We emphasize an important point: the $\mathbb{Z}_2$ symmetry is a subgroup of the original flavour group. Therefore, the stability of DM is closely related to the flavour structure of the model.
    We elaborate more on this scenario in the following. 
    \item $|Q_\chi|>2\,$: The residual symmetry is $\mathbb{Z}_{|Q_\chi|>2}$. Depending on the charge assignments, one or more particles could be stable, so that it is likely to identify one (or more) DM candidate(s). 
    
\end{itemize}
{In the following, we focus on the simplest scenario, namely $|Q_\chi|=2$, corresponding to a residual $\mathbb{Z}_2$ symmetry, while we leave a more systematic classification of stabilizing symmetries for higher charges as a future direction. 
Without loss of generality, we choose $n=1$ and $m=0$ in Eq.~\eqref{eq:beta} and Eq.~\eqref{eq:alphas} to describe the action of the nontrivial generator of the residual $\mathbb{Z}_2$. All odd values of $n$ select the same generator, whereas even values select the identity. Changing $m$ only shifts $\alpha_3$ by multiples of $4\pi$ and leaves the $\SUf{2}$ transformation unchanged.
\begin{table}[t]
\centering
\begin{tabular}{ccc|c}
\hline
 %$\SUf{2}$ representation 
 $\mathbf{R=2J+1}$  & 
 $\mathbf{Q}$ & $\mathbf{|Q\pm Q_\phi|}$ & $\mathbb{Z}_2$  \\ \hline\hline 
 \text{integer} $J$ & \text{even} & $/$ & $+$\\
 \text{integer} $J$ &\text{odd} & $/$ & $-$ \\
 \text{half-integer} $J$ & $/$ & \text{even} &  $+$\\
 \text{half-integer} $J$ & $/$ & \text{odd}& $-$ \\
\hline
\end{tabular}
\caption{Charge assignment under the residual $\mathbb{Z}_2$ symmetry for $|Q_\chi|=2$ and  generic integer $|Q_\phi|$. The symbol $/$ means that the value is irrelevant for the determination of the final $\mathbb{Z}_2$ charge.}
\label{tab:Z2chargesgeneral}
\end{table}
We now consider the action of the residual symmetry on the generic field in Eq.~\eqref{eq:generalU2}. The field can be written as an $\SUf{2}$ multiplet of dimension $R=2J+1$, namely  $\psi=(\psi_1,\dots,\psi_R)$. The $j$-th component of the multiplet transforms under the residual $\mathbb{Z}_2$ symmetry as 

\begin{equation}
    \psi_j\to e^{i\pi(Q-2\lambda_j\,Q_\phi)}\psi_j\,,
\end{equation}
where $\lambda_j$ is the $j$-th eigenvalue of the generator $T_R^3$, which can always be taken as a diagonal $R\times R$ matrix. We distinguish two possibilities: for integer representations of $\SUf{2}$, (singlet, triplet, quintuplet,...), the eigenvalues are integer numbers $\lambda_j=\pm\,(0,1,2,...)$. Therefore, every component of the multiplet transforms as $\psi_j\to \exp[i\pi Q]\,\psi_j$, so that fields with an even/odd charge $Q$ are even/odd under the $\mathbb{Z}_2$, irrespectively of the value of $Q_\phi$.

On the contrary, for half-integer representations of $\SUf{2}$ (doublet, quadruplet,...), the eigenvalues are of the form $\lambda_j=(2k_j+1)/2=\pm\,(1/2,3/2,5/2...)$, and every component of the multiplet transforms as $\psi_j\to \exp[i\pi (Q-(2k_j+1)Q_\phi)]\,\psi_j$. Therefore, if $Q$ and $Q_\phi$ have the same parity, both even or both odd numbers, the field is even under the residual  $\mathbb{Z}_2$; if $Q$ and $Q_\phi$ have opposite parity, one odd and one even, the field is odd under the $\mathbb{Z}_2$. The general transformation rules are summarized in Table~\ref{tab:Z2chargesgeneral}.  
\begin{table}[t]
\centering
\begin{tabular}{cc|c}
\hline
$\mathbf{R=2J+1}$  & 
$\mathbf{Q}$ & $\mathbb{Z}_2\sim(-1)^{Q+2J}$  \\ \hline\hline 
 \text{integer} $J$ & \text{even} & $+$\\
 \text{integer} $J$ &\text{odd}  & $-$ \\
 \text{half-integer} $J$ & \text{even} &  $-$\\
 \text{half-integer} $J$ & \text{odd}& $+$ \\
\hline
\end{tabular}
\caption{Charge assignment under the residual $\mathbb{Z}_2$ symmetry for $|Q_\chi|=2$ and $|Q_\phi|=1$.}
\label{tab:Z2charges}
\end{table}
In the following, we focus on the specific $|Q_\phi|=1$ case, whose corresponding transformation rules can be summarized as $\psi\to (-1)^{Q+2J}\,\psi$, as reported in Table~\ref{tab:Z2charges}.
The lightest state odd under $\mathbb{Z}_2$ is automatically stable and can be identified as the DM candidate. An intriguing possibility is to identify such state with a sterile neutrino, singlet of $\SUf{2}$ and the SM gauge group, with charge $Q_{N_3}$ under $\Uf{1}$, which we generically denote as $N_3$. This is automatically stable if

\begin{itemize}
    \item $Q_{N_3}$ is odd, so that $N_3\to-N_3$;
    \item all the other fields in integer representations of $\SUf{2}$, including any additional  singlet, have even charges;
    \item all the fields in half-integer representations of $\SUf{2}$, including all the doublets, have odd charges.
\end{itemize}
We emphasize once again the crucial property of this construction: the stability of DM is determined by the charges of the fields under the flavour group. The same charges fix completely the fermion masses and mixings  and are therefore constrained by their experimental values. In other words, DM stability and flavour structure are fundamentally interconnected.
This differentiates our scenario from constructions in which an external, ad hoc symmetry, completely unrelated to flavour, is imposed by hand.

}

%%%%%%%%%%%%%%%%%%%%%%%%%%%%%%%
\section{A Prototype Model with Right-Handed Neutrinos} \label{sec:prototype}
%%%%%%%%%%%%%%%%%%%%%%%%%%%%%%%

Building on the $\SUf{2}\otimes\Uf{1}$ framework of Ref.~\cite{Linster:2018avp}, we adopt a FN model extended solely by three right-handed neutrinos, $N_{1,2,3}$. This setup offers a remarkably economical framework that addresses two open problems simultaneously without introducing extra exotic matter fields: while the heavy states mediate a type-I seesaw mechanism to generate light neutrino masses, the lightest state, $N_3$, serves as a viable DM candidate whose stability is naturally guaranteed by the residual flavour symmetry of our setup.

In the original construction of Ref.~\cite{Linster:2018avp}, the right-handed neutrinos were arranged as a $\SUf{2}$ doublet, $N_a$, and a singlet, $N_3$. While phenomenologically viable, reproducing the observed neutrino mass-squared differences without fine-tuning required mapping the continuous symmetry onto an isomorphic discrete subgroup. 
This issue is circumvented by assigning all three right-handed neutrinos as $\SUf{2}$ singlets~\cite{Giarnetti:2025idu}. 

We adopt this three-singlet realization as our benchmark model; its field content and charge assignments are collected in Table~\ref{tab:scenarioN}.
{We emphasize that this construction is intended as a representative model illustrating the connection between residual DM stabilization and flavour structure. We leave a systematic analysis of realistic flavour models for future work.}

\begin{table}[t]
\centering
\begin{tabular}{c|c|c}
\hline
\textbf{Fields $\mathbf{(a=1,\,2)}$} & $\mathbf{\textbf{SU}(2)_\textbf{F}}$ & $\mathbf{\textbf{U}(1)_\textbf{F}}$   \\ \hline 
$L_a,\, E_a  $& 2 &  1 \\
$L_3$ & 1 &$ 2n_L$  \\
$E_3$ & 1 & $2n_L-2$\\
\hline
$N_1$ & 1 & $2n_1$ \\
$N_2$ & 1 & $2n_2$ \\
\rowcolor{gray!15}
$N_3$ & 1 &  $Q_{N_3}\equiv 2n_3+1$ \\
\hline
\hline
$\phi_a$ & 2 & $-1$ \\
$\chi$ & 1 & $-2$ \\
\hline
\end{tabular}
\caption{Here $n_{L,i}$ ($i=\{1,2,3\}$) are generic integer numbers.}\label{tab:scenarioN}
\end{table}

%%%%%%%%%%%%%%%%%%%%%%%%%%%%%%%%%%%
\subsection{The charged lepton sector}
\label{sec:chargedleptons}
%%%%%%%%%%%%%%%%%%%%%%%%%%%%%%%%%%

The charged lepton sector is described by
\begin{equation}
\begin{split}
    \mathcal{L}_{\rm CL}=&\frac{e_{11}}{\Lambda^4}\chi^2(\tilde{\phi}L)_1(\tilde{\phi}E)_1 H+\frac{e_{12}}{\Lambda}\chi(LE)_1 H+\frac{e_{22}}{\Lambda^2}({\phi}L)_1({\phi}E)_1 H \\
   &+e_{13}\frac{\chi^{|n_L|}}{\Lambda^{1+|n_L|}}(\tilde{\phi}L)_1E_3H+e_{23}\frac{\chi^{|n_L-1|}}{\Lambda^{1+|n_L-1|}}(\phi L)_1E_3H  \\
   &e_{31}\frac{\chi^{|1+n_L|}}{\Lambda^{1+|1+n_L|}}L_3(\tilde{\phi}E)_1H+e_{32}\frac{\chi^{|n_L|}}{\Lambda^{1+|n_L|}}L_3(\phi E)_1H+e_{33}\frac{\chi^{|2n_L-1|}}{\Lambda^{|2n_L-1|}}L_3E_3H+{\rm h.c.}\,,  
\end{split}
\label{eq:FNlep}
\end{equation}
where $L,E$ are the SM leptons and $H$ is the Higgs field. Notice that we adopt the notation in which all the fermion fields are left-handed. Specializing to the case $n_L=1$,
the Yukawa matrix of the charged sector reads

\begin{equation}
  Y_e= y_0\left( \begin{array}{ccc}
               e_{11}\epsilon_\phi^2\epsilon_\chi^2  & e_{12}\epsilon_\chi & e_{13}\epsilon_\phi\epsilon_\chi \\
      -e_{12}\epsilon_\chi   &  e_{22}\epsilon_\phi^2 & e_{23}\epsilon_\phi \\ 
      e_{31}\epsilon_\phi\epsilon_\chi^2 &  e_{32}\epsilon_\phi\epsilon_\chi & e_{33}\epsilon_\chi \\
    \end{array}
    \right)_\text{LR}\,.
\end{equation}
where the overall $y_0$ is fixed by requiring the largest charged lepton mass eigenvalue to reproduce the observed tau mass.
Without loss of generality\footnote{Furthermore, we have numerically tested that a hierarchy between the VEVs of the two flavons makes it very difficult to reproduce the PMNS matrix, unless a strong tuning among the parameters is invoked.} we can assume $\epsilon_\phi \sim \epsilon_\chi\sim\varepsilon$ and, for generic $\mathcal{O}(1)$ coefficients, the Yukawa matrix yields three charged lepton masses of the same parametric order, despite the different suppressions induced by the flavour symmetry.
A hierarchical charged lepton spectrum can instead be obtained by taking\footnote{Notice that, for $\varepsilon\simeq 0.2$, this is not a real tuning, as it can be absorbed into the Yukawa coefficients.} $e_{12}=\varepsilon^{2\div 3}e^\prime_{12}$ and $e_{13}=\varepsilon^{2\div 3}e^\prime_{13}$, with $(e^\prime_{12},e^\prime_{13})$ being $\mathcal{O}(1)$ parameters. In this way, the first row and the first column of the matrix $Y_e\,Y_e^\dagger$ can be suppressed, thus decoupling the electron from the $\mu$--$\tau$ pair. Specifically, the electron squared-mass\footnote{It is worth noticing that the same result can be obtained by imposing a hierarchy between the VEVs of the flavons, namely $\epsilon_\chi/\epsilon_\phi\sim \mathcal{O}(10^{-1})$. However, within this choice it is challenging to reproduce the mixing angles and therefore we do not consider such a possibility.} turns out to be $\mathcal{O}(\varepsilon^8)$.

Focusing on the $\mu$--$\tau$ sub-block, namely the minor $\left[Y_e\,Y_e^\dagger\right]_{23}$, we observe that
\begin{equation}
    \text{Tr}\left(\left[Y_e\,Y_e^\dagger\right]_{23}\right)=m_\mu^2+m_\tau^2\approx \mathcal{O}(\varepsilon^2)\,,\quad \text{Det}\left( \left[Y_e\,Y_e^\dagger\right]_{23}\right)=m_\mu^2m_\tau^2\approx \mathcal{O}(\varepsilon^6)\,,
\end{equation}
Thus, it follows that $m_\mu^2 \sim \mathcal{O}(\varepsilon^4)$, which implies that the hierarchy $m_\mu / m_\tau \approx \mathcal{O}(\varepsilon)$ arises naturally. Taking $\varepsilon \sim 0.2$ and free parameters in the range $[\varepsilon, 1/\varepsilon]$, it is straightforward to reproduce the hierarchy reported in the PDG~\cite{ParticleData:2026uvj}.

The charged lepton sector is thus characterized at Leading Order (LO) by a single-sector rotation within the $\mu$--$\tau$ plane only. In other words, the charged lepton rotation matrix can be written simply as
\begin{equation}
  \Ul^{(23)} = R_{23}(\thetal_{23}) =
  \begin{pmatrix}
    1 & 0 & 0 \\
    0 & \cos\thetal_{23} & \sin\thetal_{23} \\
    0 & -\sin\thetal_{23} & \cos\thetal_{23}
  \end{pmatrix}\,,\quad\text{with}\quad \tan(2\thetal_{23})=\dfrac{2\,e_{23}e_{22}}{e_{33}^2-e_{23}^2}\,.
  \label{eq:Ul_23}
\end{equation}
The (2,3) rotation is typical of models with $\mu$--$\tau$
symmetry~\cite{Arcadi:2022ojj,Ma:2001mr,Mohapatra:2005ra}, in which a maximal atmospheric angle $\thetanu_{23} = \pi/4$ and $\thetanu_{13} = 0$ in the neutrino
sector occur.

%%%%%%%%%%%%%%%%%%%%%%%%%%%%%%%%%%%
\subsection{The neutrino sector}
\label{sec:neutrinos}
%%%%%%%%%%%%%%%%%%%%%%%%%%%%%%%%%%%

The interactions among the SM leptons and the sterile neutrinos are given by

\begin{equation}
\begin{split}
    \mathcal{L}_{\rm DIR}&=\sum_{i=1,2}y_{1i}\frac{\chi^{|1+n_i|}}{\Lambda^{1+|1+n_i|}}(\tilde{\phi}L)N_i H+\sum_{i=1,2}y_{2i}\frac{\chi^{|n_i|}}{\Lambda^{1+|n_i|}}({\phi}L)N_i H \\
&+\sum_{i=1,2}y_{3i}\frac{\chi^{|n_L+n_i|}}{\Lambda^{|n_L+n_i|}}L_3N_i H\, + \text{h.c.}.
\end{split}
\label{eq:FNneut}
\end{equation}
As expected, the residual $\mathbb{Z}_2$ symmetry forbids any interaction among the SM leptons and the DM candidate $N_3$. Finally, the Majorana mass terms are

\begin{equation}\label{eq:MajoranaLag}
\begin{split}
    \mathcal{L}_{\rm M}=\Lambda\left[ \sum_{i=1,2}k_{ii}\frac{\chi^{|2n_i|}}{\Lambda^{|2n_i|}}N_iN_i+k_{12}\frac{\chi^{|n_1+n_2|}}{\Lambda^{|n_1+n_2|}}N_1N_2+k_{33}\frac{\chi^{|2n_3+1|}}{\Lambda^{|2n_3+1|}}N_3N_3\right]\,.
\end{split}
\end{equation}
The Yukawa and Majorana matrices are derived directly from the previous Lagrangians, and specializing to $n_L=n_1=n_2=1$, 
\begin{equation}\label{eq:diracMajoranaNeutrinos}
Y_\nu = \begin{pmatrix}
y_{11} \epsilon_\phi \epsilon_\chi^2 & y_{12} \epsilon_\phi \epsilon_\chi^2 & 0 \\
y_{21} \epsilon_\phi \epsilon_\chi   & y_{22} \epsilon_\phi \epsilon_\chi   & 0 \\ 
y_{31} \epsilon_\chi^2               & y_{32} \epsilon_\chi^2               & 0
\end{pmatrix}_{\!\mathrm{LR}}, \qquad 
M_N = \Lambda\begin{pmatrix}
k_{11} \epsilon_\chi^2 & k_{12} \epsilon_\chi^2 & 0 \\
k_{12} \epsilon_\chi^2 & k_{22} \epsilon_\chi^2 & 0 \\ 
0                      & 0                      & k_{33} \epsilon_\chi^{2n_3+1}
\end{pmatrix}_{\!\mathrm{RR}}.
\end{equation}
where we kept $n_3$ free, since $N_3$ does not contribute in the minimal seesaw.

The active neutrino mass matrix $m_\nu$ is obtained via the type-I seesaw mechanism, according to the master formula:

\begin{equation}
    m_\nu=-v^2\,Y_\nu\,M_N^{-1}\,Y_\nu^T\,,
     \label{eq:mnuMaiorana}
\end{equation}
where $v\simeq174$ GeV is the VEV of the Higgs field.
The first sharp prediction of our model is that one active neutrino is massless. This feature manifests directly from $\det Y_\nu = 0$ and stems from the underlying fermion–flavon interactions—specifically, the charge assignments in Table~\ref{tab:scenarioN}. These charges, in turn, are dictated by the requirement to reproduce low-energy lepton observables while simultaneously providing a viable fermionic DM candidate.

Furthermore, the neutrino mass matrix in the 2–3 block is anarchic, with eigenvalues of $\mathcal{O}(\varepsilon^2)$, where, as for the charged leptons, we consider the same symmetry breaking parameter $\varepsilon$. This might suggest Inverted Ordering (\textbf{IO}) as the most natural setup for the model, with $m_3 = 0$ and the two massive eigenstates degenerate (with the degeneracy broken at higher orders). However, since we have a minimal seesaw, we can readily identify the eigenvector corresponding to the null eigenstate as the cross product of the two columns of the Yukawa matrix, namely
\begin{equation}
    \vec{v}_1=\dfrac{1}{N}\begin{pmatrix}
        (y_{21}y_{32}-y_{22}y_{31})\\
        \varepsilon(y_{12}y_{31}-y_{11}y_{32})\\
        \varepsilon(y_{11}y_{22}-y_{12}y_{21})
    \end{pmatrix}\,,
\end{equation}
with $N$ being a normalisation. 
In \textbf{IO}, this eigenvector would form the third column of $U_\nu$, leading to $\sin^2\theta_{13}^\nu = 1-\mathcal{O}(\varepsilon^2)$. It should be noted that this is incompatible with the observed value of $\theta_{13}$, even after accounting for corrections coming from the (2,3) rotation in the charged lepton sector, as shown in Ref.~\cite{Giarnetti:2026nbd}.

Therefore, the most natural choice is to consider a Normal Ordering (\textbf{NO}), with $m_1 = 0$. 
For simplicity, it is useful to parametrize the neutrino mass matrix in Eq.~\eqref{eq:mnuMaiorana}  in terms of six real parameters\footnote{These parameters are generically complex and can therefore provide sources of CP violation, as already discussed within the same $\SUf{2}\times \Uf{1}$ framework~\cite{Linster:2018avp,Falkowski:2015zwa,Giarnetti:2025idu}. For the analysis considered here, however, we restrict ourselves for simplicity to the CP-conserving limit in the lepton sector. We stress that, once this assumption is relaxed, the phases entering the Yukawa couplings provide a natural source of CP violation.} $(x, y, z, u, w, q)$, namely
\begin{equation}
    m_\nu = m_0\begin{pmatrix}
        q\, \varepsilon^2 & u\, \varepsilon & w\, \varepsilon \\
        u\, \varepsilon & x\, & z \\
        w\, \varepsilon & z& y
    \end{pmatrix},
    \label{eq:FmnuR23}
\end{equation}
with $m_0=-\varepsilon^2 v^2/\Lambda$ being the overall mass scale. The relation among these effective parameters with the Yukawa couplings and the Majorana mass terms, are reported in Appendix~\ref{app:map}. The  neutrino mass matrix depends on 5 independent parameters, since one degree of freedom can be eliminated by exploiting the rank-2 condition. Following the standard parametrization of the neutrino mixing matrix $U_\nu=R_{23}(\thetanu_{23})R_{13}(\thetanu_{13})R_{12}(\thetanu_{12})$, we can diagonalize $m_\nu$ sequentially. We first perform a rotation in the $(2,3)$ plane with an angle $\theta_{23}^\nu$ chosen to diagonalize the leading $\mathcal{O}(\epsilon^2)$ sub-block. This yields
\begin{equation}
    R_{23}^T(\theta_{23}^\nu) m_\nu R_{23}(\theta_{23}^\nu) = m_0\begin{pmatrix}
        q^\prime \epsilon^2 & u^\prime\epsilon & w^\prime \epsilon \\
        u^\prime \epsilon & \lambda_2 & 0 \\
        w^\prime \epsilon & 0 & \lambda_3
    \end{pmatrix}\,,\quad\text{with}\quad \tan(2\thetanu_{23})=\dfrac{2\,z}{y-x}\,,
    \label{eq:mnuR23}
\end{equation}
where $q^\prime, u^\prime, w^\prime$ are redefined independent $\mathcal{O}(1)$ coefficients. The rotation in the $(2,3)$ plane yields $\tan 2\theta_{23}^\nu \sim \mathcal{O}(1)$, showing that $\theta_{23}^\nu$ directly depends on the $\mathcal{O}(1)$ coefficients of the Lagrangian. Consequently, a large or near-maximal mixing angle naturally emerges.

Note that $m_0 \lambda_{(2,3)}$ do not yet represent the final active neutrino masses, as they will receive sub-leading shifts from the subsequent rotations in the $(1,3)$ and $(1,2)$ sectors. Since $m_1=0$ by construction, we do not expect large deviation, and we can safely identify the active neutrino masses with $m_{(2,3)}=m_0\lambda_{(2,3)}$.
We can directly identify the scaling behavior of the mixing angles $\thetanu_{12}$ and $\thetanu_{13}$ from the rotated matrix $m_\nu^\prime$, namely
\begin{equation}
    \tan 2\thetanu_{12} \sim \dfrac{u^\prime \varepsilon}{\lambda_2}\,, \quad \tan 2\thetanu_{13} \sim \dfrac{w^\prime \varepsilon}{\lambda_3}\,,
    \label{eq:scaling_angles}
\end{equation}
A few comments are in order. If $x y (1+\epsilon) \approx z^2$ the determinant of the sub-block (2,3) of $m_\nu$ is zero up to $\mathcal{O}(\varepsilon)$, so that $m_2\sim \epsilon\,m_3$ and the correct value of $\Delta m^2_\mathrm{sol}/|\Delta m^2_\mathrm{atm}|=m_2^2/m_3^2\sim\mathcal{O}(\varepsilon^2)$ is obtained. The rank-two condition, $ \text{Det}[m_\nu^\prime]=0$, together with $m_2\sim\varepsilon m_3$, implies $u^\prime\sim\sqrt{\varepsilon}$, yielding $\tan 2\thetanu_{12}\sim\sqrt{\varepsilon}$. This result in $\thetanu_{12}$ that can be (nearly) maximal provided that all the parameters are of order one, and $\thetanu_{13}$ that is suppressed by and order $\varepsilon$.
Since the charged lepton correction is restricted strictly to a pure rotation in the (2,3) plane---which, as explicitly demonstrated in Appendix~\ref{app:PMNS}, selectively affects only the atmospheric mixing angle $\theta_{23}$---the resulting predictions for the remaining angles in the neutrino sector must be directly compared against the NuFit~\cite{Esteban:2024eli} global fit values for the solar ($\theta_{12}$) and reactor ($\theta_{13}$) angles. 
From an analytical standpoint, these physical observables can be effectively reproduced by the underlying symmetry structure of the model without requiring any  ad-hoc choice of parameters. In the following section, we present a  numerical analysis to  confirm these analytical results.

%%%%%%%%%%%%%%%%%%%%%%%%%%%%%%%%%%%
\subsection{Numerical analysis and results}
\label{sec:numerical}
%%%%%%%%%%%%%%%%%%%%%%%%%%%%%%%%%%%
\begin{table}[!t]
\centering
\renewcommand{\arraystretch}{1.2}
\begin{tabular}{l c} 
\toprule
\textbf{Parameter}$\qquad\qquad$ & \textbf{Best-fit value and $\mathbf{1\sigma}$ range} \\ 
\midrule
$r \equiv \Delta m^2_\text{sol}/|\Delta m^2_\text{atm}|$ & $0.0300\pm0.0005$ \\
$\sin^2\theta_{12}$ & $0.3088^{+0.0067}_{-0.0066}$ \\
$\sin^2\theta_{13}$ & $0.02248^{+0.00055}_{-0.00059}$ \\
$\sin^2\theta_{23}$ & $0.470^{+0.017}_{-0.014}$ \\
\midrule
$m_e / m_\mu$ & $0.0048 \pm 0.0002$ \\
$m_\mu / m_\tau$ & $0.0565 \pm 0.0045$ \\ 
\bottomrule
\end{tabular}
\caption{{Neutrino observables and their 1$\sigma$ ranges, from NuFIT 6.1~\cite{Esteban:2024eli} using the dataset with SK atmospheric data~\cite{Super-Kamiokande:2017yvm}, for \textbf{NO}. The mass ratios of charged leptons are taken from~\cite{Feruglio:2017spp}. Here we defined $\Delta m^2_\text{sol}\equiv m_2^2-m_1^2$, $|\Delta m^2_\text{atm}|\equiv |m_3^2-m_1^2|$. The running of low-energy neutrino observables can be neglected since in our scan we do not consider a quasi-degenerate spectrum \cite{Antusch:2003kp}.}}
\label{tab:observablesNuFit}
\end{table}
\begin{figure}[t]
    \centering
    \includegraphics[width=0.9\textwidth]{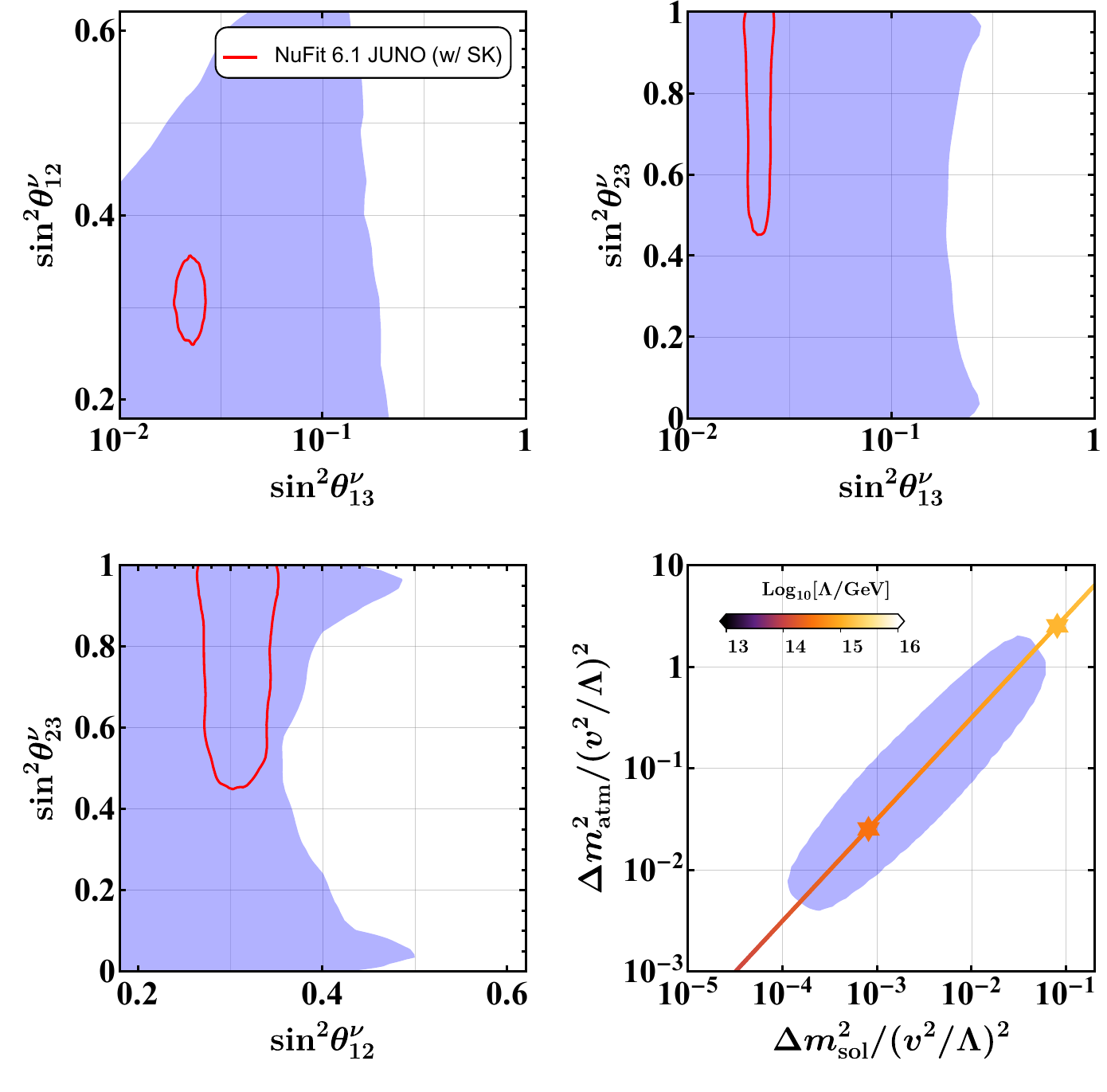}
  \caption{{
  Model predictions (blue) in the $(\sin^2\theta^\nu_{13},\sin^2\theta^\nu_{12})$ (top left panel), $(\sin^2\theta^\nu_{13},\sin^2\theta^\nu_{23})$ (top right panel) and  $(\sin^2\theta^\nu_{12},\sin^2\theta^\nu_{23})$ (bottom left panel) planes compared  with the $90\%$  contours obtained varying the PMNS parameters (red contours), see text for details. The overlap between the blue and red regions shows that the observed PMNS mixing angles arise naturally from the model without fine-tuning. In the bottom right panel the predicted neutrino mass-squared differences in the $(\Delta m^2_{\rm sol}/(v^2/\Lambda)^2,\;\Delta m^2_{\rm atm}/(v^2/\Lambda)^2)$ plane. The colored line corresponds to points reproducing the observed ratio $r \equiv \Delta m^2_{\rm sol}/\Delta m^2_{\rm atm}$, with the color indicating the value of $\text{Log}_{10}(\Lambda/{\rm GeV})$. Star markers denote integer values of $\text{Log}_{10}(\Lambda/{\rm GeV})$. The overlap of the $r$-compatible line with the $90\%$ region indicates that the observed neutrino mass hierarchy is naturally reproduced for a right-handed neutrino mass scale $\Lambda \sim \mathcal{O}(10^{14})\,{\rm GeV}$, while significantly smaller or larger values are disfavored within the present framework.}
}
    \label{fig:neutrinomixingplanes}
\end{figure}
Models based on continuous flavour symmetries typically feature a large number of parameters, with the dimensionality of the parameter space exceeding the number of low-energy observables, i.e., the three mixing angles $(\theta_{12},\theta_{13},\theta_{23})$ and\footnote{We refer the reader to Appendix~\ref{app:PMNS} for notations and numerical parameters of the mixing matrix $U_\mathrm{PMNS}$.} the three mass ratios $(m_e/m_\mu, m_\mu/m_\tau, r\equiv\Delta m^2_\mathrm{sol}/|\Delta m^2_\mathrm{atm}|)$. The same applies to our representative model, which features $8$ and $9$ real parameters\footnote{The quoted numbers refer to the CP-conserving limit adopted in our analysis. If the parameters are allowed to be generically complex, each independent coefficient carries both a modulus and a phase, thereby naively doubling the number of real parameters. The actual number of physical degrees of freedom is smaller, since a subset of these phases can be removed through right-handed fields rephasing.} stemming from the charged lepton and neutrino sectors, respectively. 
Thus, to verify that our model is able to reproduce the dimensionless experimental values in Table~\ref{tab:observablesNuFit}, we perform an unconstrained scan over the free real couplings within $[\varepsilon, 1/\varepsilon]$ and present the  regions containing the $90\%$ of neutrino mixing angles predicted by the model in the $(\sin^2\theta_{13}^\nu, \sin^2\theta_{12}^\nu)$, $(\sin^2\theta_{13}^\nu, \sin^2\theta_{23}^\nu)$ and $(\sin^2\theta_{12}^\nu, \sin^2\theta_{23}^\nu)$ planes. The only constraint we impose during the numerical scan is requiring the ratio $r$ to lie in the range $[10^{-2}, 10^{-1}]$.

As in Ref.~\cite{Giarnetti:2026nbd}, we exploit the single-sector rotation in the charged lepton sector by scanning $\thetal_{23}\in [0^\circ, 45^\circ]$ (we restrict to the lower octant, with no loss of generality) and sampling the physical PMNS angles from Gaussian distributions\footnote{
While this approach is sufficient for illustrating the viability of the model, it does not constitute a full statistical fit. Such an analysis would require the use of the complete neutrino-oscillation likelihood, including the non-Gaussian nature of the PMNS parameter uncertainties and the correlations among the oscillation observables.} based on the NuFIT $3\sigma$ uncertainties~\cite{Esteban:2024eli}. 
In Fig.~\ref{fig:neutrinomixingplanes}, the blue regions represent the $90\%$ points generated by our model, while the red contours enclose the $90\%$  contours consistent with experimental data.
The clear overlap between the blue regions and the red contours demonstrates that the model naturally reproduces the PMNS mixing angles without requiring any fine-tuning. It should be emphasized that this study does not constitute a fit of the model. Rather, it demonstrates that, without assuming any prior on the Lagrangian parameters, the model naturally generates points that lie within the $90\%$ regions consistent with the PMNS matrix for setups characterized by a single-sector charged lepton rotation parameterized, at LO, by the single angle $\thetal_{23}$. 

As illustrated in Fig.~\ref{fig:neutrinomixingplanes}, the allowed parameter space regions represented by the red contours appear remarkably constrained. The only notable exception is observed in the $\sin^2\thetanu_{23}$ profile, which stems from the fact that the entire geometric dependence on the unconstrained charged lepton mixing angle $\thetal_{23}$ is selectively absorbed within the atmospheric sector. 
In the remaining mixing sectors, the residual theoretical uncertainties are strictly dictated by the experimental boundaries provided by the global fit. Thanks to the unprecedented precision delivered by next-generation oscillation experiments, most notably JUNO \cite{JUNO:2015zny}, the solar and reactor parameters are now exceptionally well determined.
For further details, check  the analytical expressions of $U_\nu$ in light of the single-sector-rotation in the charged lepton sector in Appendix~\ref{app:PMNS}.

The right-handed neutrino mass scale $M_N\sim \Lambda$ can be inferred from the overall scale $v^2/\Lambda$ appearing in the light neutrino mass matrix, see Eq.~\eqref{eq:mnuR23}, and is not included as an independent prediction in the numerical scan, as the observables constrained by NuFit depend only on the mass ratio $r$. Consequently, $\Lambda$ can be determined by matching the model predictions to the experimental value of $|\Delta m^2_\mathrm{atm}|$ (or $\Delta m^2_\mathrm{sol}$). Specifically, we adopt $|\Delta m^2_\mathrm{atm}| = 2.511^{+0.021}_{-0.020} \times 10^{-3}\,\text{eV}^2$ for \textbf{NO} from NuFit 6.1~\cite{Esteban:2024eli}. 
In the bottom right panel of Fig.~\ref{fig:neutrinomixingplanes}, we show the regions containing $90\%$ of the model points in the $(\Delta m^2_{\mathrm{atm}}$, $\Delta m^2_{\mathrm{sol}})$ plane. The colored line highlights the points that reproduce the correct value of the ratio $r$, with the color coding indicating the value of $\text{Log}_{10}(\Lambda/\text{GeV})$ and stars marking integer exponent values. This provides a clear estimate of the prediction for $\Lambda$: within the $90\%$ region, the right-handed neutrino mass scale when the experimental value of $r$ is matched, is $\Lambda \sim \mathcal{O}(10^{14})\,\text{GeV}$. Mass scales of lower (or higher) orders of magnitude fall outside the $90\%$ contour and are thus disfavored by the present model. It should be noted that $\Lambda$ does not directly correspond to the physical right-handed neutrino masses, as the eigenvalues of the Majorana mass matrix undergo extra suppression factors proportional to the breaking parameter $\varepsilon$, as shown in Eq.~\eqref{eq:diracMajoranaNeutrinos}.

We now proceed to discuss the phenomenological predictions of our model. First, the \textbf{NO} scenario is strongly favored over the \textbf{IO}. The lightest neutrino mass is predicted to be exactly vanishing, $m_1 = 0$, which allows us to obtain simple analytical expressions for both the effective Majorana mass in neutrinoless double-beta decay ($|m_{\beta\beta}|$) and the effective neutrino mass in tritium decay ($m_\beta^{\text{eff}}$).

As already mentioned, all parameters are assumed to be real for simplicity. Nevertheless, CP-violating phases can be naturally incorporated into the model. In a minimal seesaw, one of the Majorana phases can be eliminated, allowing us to set $\alpha_1 = 0$ (see Appendix~\ref{app:PMNS}), while the second Majorana phase, $\alpha_2$, remains physical. Owing to the reduced number of independent low-energy parameters, the rank-two structure of the minimal seesaw leads to simple analytical expressions for several neutrino observables.\footnote{These predictions are generic to minimal seesaw scenarios with \textbf{NO} and a massless neutrino. Indeed, similar results are obtained, e.g., in Refs.~\cite{Wang:2019ovr,Xing:2020ald,Marciano:2024nwm,King:2013iva}.} In particular, the effective mass governing neutrinoless double-beta decay reads
\begin{equation}\label{mbetabeta}
\left|m_{\beta\beta}\right|
=
\left|\sum_i m_i U_{ei}^2\right|
=
\sqrt{\Delta m_{\mathrm{atm}}^2}
\left|
\sqrt{r}\,s_{12}^2c_{13}^2
+
s_{13}^2
e^{i(\alpha_2-2\delta_{\rm CP})}
\right|
\sim \mathcal{O}(1)\,{\rm meV},
\end{equation}
whereas the effective neutrino mass probed in single-beta decay reads:
\begin{equation}\label{mtritium}
m_{\beta}^{\rm eff}
=
\sqrt{\sum_i m_i^2 |U_{ei}|^2}
=
\sqrt{\Delta m_{\mathrm{atm}}^2}
\sqrt{
r\,s_{12}^2c_{13}^2+s_{13}^2
}
\sim \mathcal{O}(10)\,{\rm meV},
\end{equation}
in full agreement with the results in Ref.~\cite{Esteban:2024eli}. Here, $s_{ij}\equiv \sin\theta_{ij}$ and $c_{ij}\equiv \cos\theta_{ij}$, with $\theta_{ij}$ denoting the PMNS mixing angles. Finally, the non-vanishing neutrino mass eigenstates lie in the tens of meV region, yielding a total sum of masses $\sum_i m_i \sim 0.06$ eV. This is well within the latest cosmological upper bound of $0.115\,\text{eV}$ ($95\,\%\,\text{C.L.}$)~\cite{Esteban:2024eli}. This value is primarily governed by the \textbf{NO} scenario with a vanishing lightest neutrino mass:
\begin{equation}\label{cosmo_sum_no}
\sum_{i=1}^{3}m_i
=
\sqrt{\Delta m_{\mathrm{atm}}^2}
\left(
1+\sqrt{r}
\right)
\sim 0.06~{\rm eV}.
\end{equation}
Furthermore, following Ref.~\cite{Calibbi:2025rxn,Bertolini:2020hjc}, the lower bound on the cutoff scale reaches $10^9\,\text{GeV}$ in light of the constraints induced by charged Lepton Flavor Violation (cLFV). Those limits are derived taking into account the processes induced by the pseudo-Nambu-Goldstone bosons (pNGBs) arising from $\SUf{2}\otimes\Uf{1}$ symmetry breaking. To avoid massless states in the spectrum, we postulate a soft symmetry breaking term, potentially generated by quantum gravity as in~\cite{Calibbi:2025rxn}. In light of this, the pNGBs are typically much lighter than the cut-off scale $\Lambda$.
In the limit of real parameters, the pNGBs feature only charged Lepton Flavor Conserving (cLFC) pseudoscalar couplings~\cite{Ema:2016ops,Calibbi:2016hwq,Calibbi:2020jvd,Mandal:2023jnv}.\footnote{In contrast, cLFV couplings can feature both scalar and pseudoscalar components even in the absence of CP-violating phases.} Interestingly, when complex CP-violating phases are present, they induce an additional cLFC scalar coupling for the pNGB~\cite{Bertolini:2020hjc}, leading to distinct fifth-force signatures across various mass scales.

%%%%%%%%%%%%%%%%%%%%%%%%%%%%%%%%%%%%%%%%%%
\section{Dark Matter production and constraints}
\label{sec:DMphenomenology}
%%%%%%%%%%%%%%%%%%%%%%%%%%%%%%%%%%%%%%%%%%
In this section, we assess the viability of the sterile neutrino $N_3$ as a DM candidate.  As emphasized in the previous sections, $N_3$ is stable due to the accidental residual flavour $\mathbb{Z}_2$ symmetry.
Here, we track the evolution of its number density in the early Universe and require the resulting relic abundance to match the value measured by the Planck collaboration, $\Omega_{\rm DM}h^2=0.1200\pm0.0012$~\cite{Planck:2018vyg}, with $\Omega_{\rm DM}$ denoting the DM energy density normalized to the critical density and $h\simeq0.67$ the reduced Hubble parameter. Furthermore, we make sure that the interactions of $N_3$ satisfy all
experimental and observational constraints. In the following discussion, we refer to $N_3$ as the DM candidate. 

The DM cosmological history depends crucially  on the strength of its interactions with the SM particles. In our framework, these arise from higher-dimensional operators and are therefore suppressed by the UV cut-off scale $\Lambda$. As discussed in the previous section, 
constraints from cLFV processes roughly require $\Lambda\gtrsim v_{\phi,\chi}\gtrsim 10^{9}$ GeV~\cite{Calibbi:2025rxn}. Moreover, as shown in Fig.~\ref{fig:neutrinomixingplanes}, scales as high as $\Lambda\sim 10^{14}$ GeV are typically needed to reproduce viable active neutrino masses, in the absence of suppressed Yukawa couplings.
The interactions of $N_3$ with the SM fields arise either from higher-dimensional operators suppressed by the cutoff scale $\Lambda$ or through flavon-mediated processes, with couplings controlled by the flavour symmetry breaking scales $v_{\phi,\chi}$. For the high scales considered here and reheating temperatures considered below, these interactions are sufficiently suppressed that $N_3$ never reaches thermal equilibrium with the SM bath.
Moreover, motivated by the discussion in the previous section, we assume that the pNGBs are much lighter than the DM, potentially with masses in the keV-to-MeV range, as considered in~\cite{Calibbi:2025rxn}. For the values of $v_{\chi,\phi}$ considered in our analysis
the pNGBs do not thermalize with the SM bath (see also Ref.~\cite{Mandal:2023jnv}). 

\begin{figure}[t]
    \centering
\includegraphics[width=0.8\textwidth]{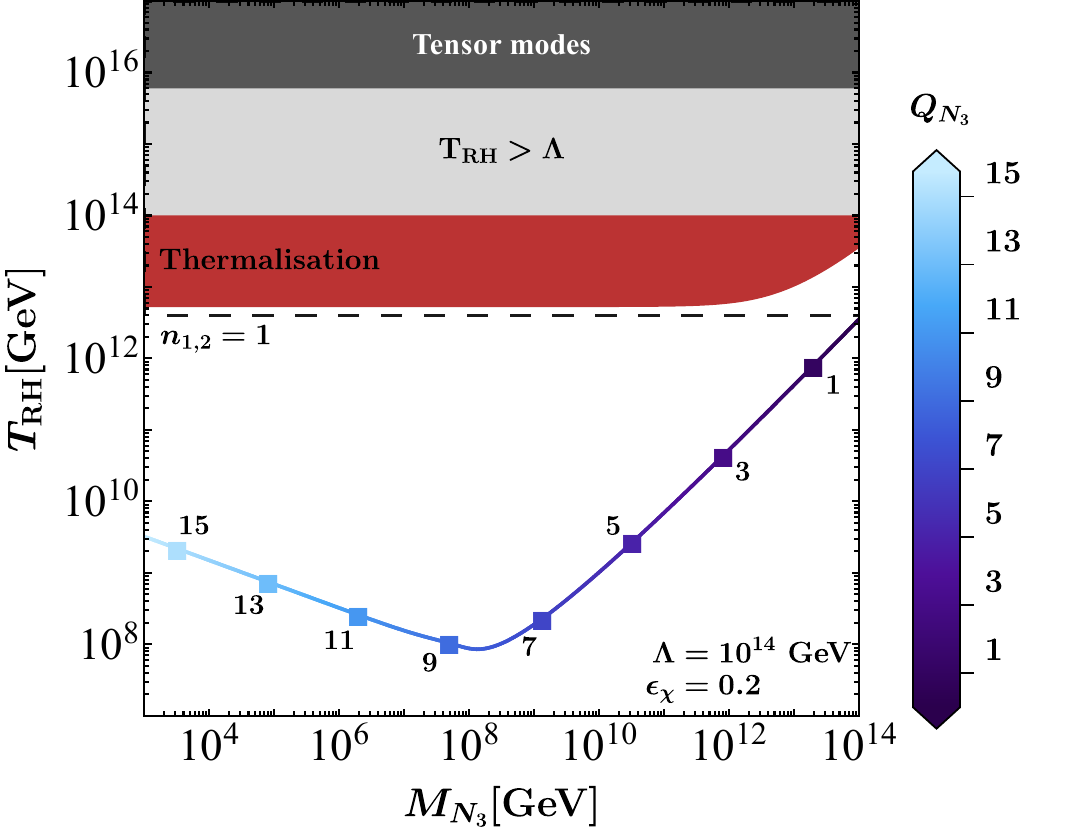}
   \caption{The DM relic abundance is reproduced along the colored line for the benchmark $\Lambda=10^{14}$ GeV and $\epsilon_\chi=0.2$. The color legend indicates the corresponding DM $\Uf{1}$ charge, with the squares denoting the physical values (odd integers). The dark gray region is excluded by tensor inhomogeneities, while in the light gray region the EFT approach is not valid. The dashed line corresponds to the value of the masses of the additional sterile neutrinos $N_{1,2}$. }
    \label{fig:DMfinal}
\end{figure}

In the absence of a thermal DM population, and assuming a negligible abundance right after the end of inflation, $N_3$ particles are produced from  occasional annihilations of SM particles in the thermal bath, a  mechanism known as 
\textit{freeze-in}~\cite{Hall:2009bx}.
Following the standard notation, the Hubble expansion rate (assuming standard radiation-dominated universe) and the entropy density are defined as 

\begin{equation}
H(T)=\sqrt{\frac{4\pi^3 g_*(T)}{45}}\frac{T^2}{M_{\rm Pl}} \, , \qquad s(T)=\frac{2\pi^2}{45}g_s(T)T^3 \, ,
\end{equation} 
where $T$ is the temperature of the SM thermal bath, and the functions $g_s$ and $g_*$ count the number of relativistic degrees of freedom in the thermal equilibrium ($g_s=g_*$ for $T\gtrsim\text{MeV}$).
In light of the exact $\mathbb{Z}_2$ symmetry, the leading operators for DM production contain two powers of $N_3$. Let us consider a generic operator of the form
\begin{equation}\label{eq:generalOp}
    c_\mathcal{O}\frac{\mathcal{O}_{\rm SM}\mathcal{O}_{N_3}}{\Lambda^p}\,,
\end{equation}
where  $\mathcal{O}_{\rm SM}$ is an operator, at least quadratic, made by SM fields, $\mathcal{O}_{N_3}$ is a bilinear in the DM field, either $N_3N_3$ or $N_3^\dagger\overline{\sigma}_\mu N_3$, and $p\geq1$ is the appropriate power. This induces annihilation processes $\sum_{i}{\rm SM}_i\to N_3N_3$ (or ${\sum_{i}\rm SM}_i\to N_3N_3\,\sum_j{\rm SM}'_j$), where the initial state includes two or more SM particles. In the relativistic limit, the corresponding interaction rate density scales as $\gamma_{\rm eq}\sim c_\mathcal{O}^2T^{2(p+2)}/\Lambda^{2p}$. The DM abundance is expressed in terms of $Y=n/s$, being $n$ its number density. We introduce the reheating temperature $T_{\rm RH}$ as the highest temperature of the thermal bath after inflation (in the instantaneous reheating approximation) and define $z=T_{\rm RH}/T$. Assuming a negligible initial DM abundance and working in the freeze-in regime, $Y\ll Y_{\rm eq}$, where $Y_{\rm eq}\equiv n_{N_3}^{\rm eq}/s$ denotes the equilibrium yield of $N_3$, the relevant Boltzmann equation for the evolution of the DM abundance reads $sHz\,dY/dz=2\gamma_{\rm eq}$. This is solved by
\begin{equation}\label{eq:generalY}
    Y=\frac{2}{2p-1}\frac{\gamma_{\rm eq}}{Hs}\bigg|_{T=T_{\rm RH}}\,.
\end{equation}
The observed DM abundance is then reproduced if $Y=T_{\rm eq}/M_{N_3}\simeq0.4\,{\rm eV}/M_{N_3}$.
The freeze-in mechanism is sensitive to UV physics when driven by non-renormalizable operators~\cite{Elahi:2014fsa}. As shown by Eq.~\eqref{eq:generalY}, the DM production is maximally efficient at the highest
available energy scales, i.e $T_{\rm RH}$. Thus, whenever $\Lambda<T_{\rm RH}$, the details of the UV completion of the model are needed to compute the DM abundance. To  avoid this, we restrict our analysis to values of the cut-off and symmetry breaking scale satisfying
\begin{equation}
   \Lambda>v_{\phi}\sim v_{\chi}\gg T_{\rm RH}\gg v\,.
\end{equation}
Within this assumption, the $\SUf{2}\otimes \Uf{1}$ flavour symmetry is spontaneously broken before the end of inflation and never restored afterwards.
The lower dimensional operators of the form of Eq.~\eqref{eq:generalOp},
compatible with the $\SUf{2}\otimes \Uf{1}$ symmetry arise at dimension 6 as 
\begin{equation}\label{eq:Op6}
  \frac{1}{\Lambda^2} ( N_3^\dagger\overline{\sigma}^\mu N_3)\left[c_f\,f^\dagger\overline{\sigma}_\mu f+c_H H^\dagger i \overleftrightarrow D_\mu\, H\right]\,,
\end{equation}
where $f=\{Q,U,D,L,E\}$ are the SM fermions and $c_i$ unknown coefficients. Assuming that all the coefficients are of $\mathcal{O}(1)$ and taking into account the multiplicities, the total contribution of $f^\dagger f \,(H^\dagger H)\to N_3N_3$ to the DM abundance  parametrically scales as 
\begin{equation}\label{eq:Y}
    Y\sim  10^{-4}\frac{M_{\rm Pl}\max\{T_{\rm RH}^3,M_{N_3}^3\}}{\Lambda^4}e^{-2M_{N_3}/T_{\rm RH}}\,,
\end{equation} 
where the exponential factor takes into account that DM particles heavier than the reheating temperature are only produced in the tail of the number density distribution of bath particles. This result must be understood as a parametric estimate, while  a precise analysis of the coefficients in concrete UV completions is beyond the scope of the current discussion.
Given that the interaction rate of the processes induced by the operators in Eq.~\eqref{eq:Op6} scales as $\Gamma\sim T^5/\Lambda^4$, $N_3$ never thermalized with the SM bath as long as $T_{\rm RH}\lesssim 10^{13}$ GeV.
Notice that the effective operator 

\begin{equation}\label{eq:HiggsOp}    \widetilde{c}_H\frac{\chi^{Q_{N_3}}}{\Lambda^{1+Q_{N_3}}}(H^\dagger H)N_3N_3\,,
\end{equation}
can also be generated in the UV theory. When the flavon acquires VEV, this reduces to an effective dimension-5 operator, with an extra $(M_{N_3}/\Lambda$) suppression,

\begin{equation}\label{eq:HiggsDM}
\widetilde{c}_H\epsilon_\chi^{Q_{N_3}}\frac{(H^\dagger H)N_3^2}{\Lambda}=\frac{\widetilde{c}_H}{k_{33}}\,\frac{M_{N_3}}{\Lambda}\frac{(H^\dagger H)N_3N_3}{\Lambda}\,.
\end{equation} 
The resulting contribution to the DM cosmological abundance is sub-leading for $M_{N_3}\ll T_{\rm RH}$, while, in the opposite regime, it is comparable to that in Eq.~\eqref{eq:Y}, providing $\mathcal{O}(1)$ corrections, which can be absorbed in the UV coefficients\footnote{The operator in Eq.~\eqref{eq:HiggsOp} is also generated within the model by integrating the heavy flavon field. However, such interactions arise with a coefficient proportional to $\lambda_{H\chi}$, the quartic coupling associated to the term $\lambda_{H\chi}(H^\dagger H)(\chi^\dagger\chi)$ in the scalar potential. This interaction induces a correction to the quadratic Higgs coupling, which need to be tuned by $\mu_H^2$ to get the physical weak scale, so that $\lambda_{H\chi}\lesssim  \lambda_H(v/v_\chi)^2$ is negligibly small.}. We also checked that the annihilation processes mediated by the off-shell exchange of the radial flavon modes and the corresponding pNGBs provide a sub-leading contribution to the total DM relic abundance.
The DM relic abundance is reproduced along the colored line  in Fig.~\ref{fig:DMfinal}, for representative values of $\Lambda=10^{14}$ GeV and $\epsilon_\chi=0.2$. At lower DM masses, the curve scales approximately as $M_{N_3}\propto T_{\rm RH}^{-3}$, whereas exponential suppression becomes important at higher masses.
The color scale indicates the corresponding charge $Q_{N_3}$ with the physical odd integer values marked by a square. In the upper gray region, $\Lambda<T_{\rm RH}$ and a reliable calculation requires the knowledge of the UV completion of the model. The bound on primordial tensor perturbations imply $T_{\rm RH}\lesssim 6\times 10^{15}$, excluding the upper black region. 

The dashed line corresponds to the value of the mass of the heavy steriles $N_{1,2}$ for a minimal choice of charges $n_{1,2}=1$ and $\mathcal{O}(1)$ parameters in Eq.~\eqref{eq:MajoranaLag}. When the reheating temperature lies well below this value, the production of $N_{1,2}$ is exponentially suppressed and their thermal population is negligible. In the red region, the interactions among $N_3$ and the SM fermions are strong enough to keep the DM particle in thermal equilibrium at large temperatures. The subsequent relativistic DM decoupling would then lead to overclosure.
Finally, the DM particle is so feebly coupled to evade safely any constraint from direct and indirect searches.
\noindent
Furthermore, light pNGBs may provide an additional cosmological signature. Although their masses are free parameters of the effective theory and we mostly focus on the keV-to-MeV range, one or more of them could be as light as $m_a\lesssim 0.1$ eV.\footnote{Such a possibility has also been considered in flavour models acting on quarks, where one of the Goldstone bosons is identified with the QCD axion~\cite{Calibbi:2016hwq,Ema:2016ops}.} These states would remain relativistic at the time of Cosmic Microwave Background (CMB) formation, potentially contributing to dark radiation, conventionally parametrized by $\Delta N_{\rm eff}$~\cite{ParticleData:2026uvj}. The current Planck bound, $\Delta N_{\rm eff}\leq 0.285$ (95\% C.L.)~\cite{Planck:2018vyg}, is expected to improve by roughly an order of magnitude in future experiments. For non-thermalized pNGBs, however, determining $\Delta N_{\rm eff}$ requires accounting for their non-thermal distributions and production mechanisms. In particular, Ref.~\cite{DEramo:2023nzt} investigates dark radiation produced through decays and $2\to 2$ scatterings of bath particles yielding a single relativistic dark particle, as can occur in our scenario through SM annihilations producing a pNGB. A quantitative analysis is left for future work.
%%%%%%%%%%%%%%%%%%%%%%%%%%%%%%%%%%%
\section{Conclusions }
\label{sec:conclusions}
%%%%%%%%%%%%%%%%%%%%%%%%%%%

In this work, we have proposed a novel, model-independent framework in which DM stability emerges naturally from residual discrete symmetries left unbroken by the spontaneous breaking of continuous flavour symmetries. Rather than introducing ad hoc stabilizing symmetries, DM protection is an inherent byproduct of the same flavour dynamics responsible for generating fermion mass hierarchies and mixing patterns.

As a concrete prototype, we embedded this mechanism in a $\SUf{2} \otimes \Uf{1}$ Froggatt-Nielsen framework complemented by three right-handed neutrino singlets. 
The spontaneous symmetry breaking yields a residual $\mathbb{Z}_2$ symmetry under which the lightest right-handed neutrino, $N_3$, is odd, ensuring its absolute stability. Simultaneously, the remaining right-handed states generate light neutrino masses via a minimal type-I seesaw mechanism. Our numerical analysis confirms that the model naturally reproduces the observed PMNS mixing angles, charged lepton mass hierarchies, and mass-squared differences without fine-tuning. The setup strongly favors Normal Ordering with a strictly vanishing lightest neutrino mass ($m_1 = 0$), yielding testable predictions for neutrinoless double-beta decay, single-beta decay, and the sum of active neutrino masses.

Regarding DM phenomenology, the suppressed interactions mediated by higher-order operators dictate that $N_3$ is produced non-thermally via the freeze-in mechanism in the early Universe. We showed that the observed relic abundance can be matched for heavy DM masses and realistic reheating temperatures while respecting cosmological, cLFV, and astrophysical constraints.

In summary, continuous flavour symmetries offer a unified framework that addresses the origin of fermion masses and mixings alongside the stability, and production of DM.

Future directions include extending this residual stabilization mechanism to larger discrete subgroups $\mathbb{Z}_N$, potentially accommodating multi-component DM scenarios~\cite{Mescia:2024rki,BasBeneito2022,YagunaZapata2020,BelangerMjallalPukhov2022}, as well as to non-Abelian symmetry groups within broader model constructions.
Gauging the underlying continuous flavor symmetries introduces additional exotic fermions necessary for anomaly cancellation, leading to testable UV completions and potentially establishing connections with matter parity and, in suitable supersymmetric embeddings, $R$-parity~\cite{Martin1992,Martin1996,HerreroGarciaLandiniYanagida2025}.
Furthermore, a comprehensive analysis of the resulting light and ultra-light pNGB phenomenology is warranted, systematically addressing flavor-induced correlations across CP-violating scalar couplings in laboratory and astrophysical searches, contributions to $\Delta N_{\mathrm{eff}}$, and fifth-force constraints.

\acknowledgments

The authors thank G.~Isidori for useful discussions
and L.~Calibbi and J.~Herrero-García for comments
on the manuscript.

SM acknowledges financial support from the
Spanish grant CPI-25-643, associated with project
CNS2022-136005, funded by
MICIU/AEI/10.13039/501100011033 and the European
Union NextGenerationEU/PRTR.
SM also thanks A.~Giarnetti for helpful discussions.

\appendix
%
%%%%%%%%%%%%%%%%%%%%%%%%%%%%%
\section{Parametrizations of the rotation matrices $\UPMNS\equiv U_l^\dagger\, U_\nu$ and   $U_\nu$}
\label{app:PMNS}
%%%%%%%%%%%%%%%%%%%%%%%%%%%%%

The lepton mixing matrix $\UPMNS\equiv U_l^\dagger\, U_\nu$, which governs neutrino flavour oscillations, stems from the mismatch between $U_l$ and $U_\nu$, the unitary matrices that diagonalize the charged-lepton and neutrino sectors, respectively~\cite{Schechter:1980gr}. Moreover, it can be parameterized in the standard PDG~\cite{ParticleData:2026uvj} representation using three Euler mixing angles $\theta_{12}$, $\theta_{23}$, and $\theta_{13}$, along with one Dirac CP-violating phase $\delta_{\text{CP}}$. Factorizing the matrix as $U_{\text{PMNS}} = R_{23}(\theta_{23}) R_{13}(\theta_{13}, \delta_{\text{CP}}) R_{12}(\theta_{12}) \mathcal{P}$, its explicit form is given by
\begin{equation}
\UPMNS  = 
\begin{pmatrix}
c_{12} c_{13} & s_{12} c_{13} & s_{13} e^{-i \delta_{\text{CP}}} \\
-s_{12} c_{23} - c_{12} s_{23} s_{13} e^{i \delta_{\text{CP}}} & c_{12} c_{23} - s_{12} s_{23} s_{13} e^{i \delta_{\text{CP}}} & s_{23} c_{13} \\
s_{12} s_{23} - c_{12} c_{23} s_{13} e^{i \delta_{\text{CP}}} & -c_{12} s_{23} - s_{12} c_{23} s_{13} e^{i \delta_{\text{CP}}} & c_{23} c_{13}
\end{pmatrix}
\cdot \mathcal{P},
\end{equation}
where $c_{ij} \equiv \cos\theta_{ij}$ and $s_{ij} \equiv \sin\theta_{ij}$. If neutrinos are of Majorana nature, the diagonal phase matrix $\mathcal{P} = \operatorname{diag}(1, e^{i\alpha_1/2}, e^{i\alpha_2/2})$ introduces two additional physical CP-violating phases $\alpha_1$ and $\alpha_2$, whereas $P = \mathbf{1}_{3\times 3}$ in the Dirac case.
The three PMNS mixing angles can be extracted from the matrix elements via
\begin{equation}
  \sin^2\theta_{13} = |U_{e3}|^2, \qquad
  \sin^2\theta_{12} = \frac{|U_{e2}|^2}{1-|U_{e3}|^2}, \qquad
  \sin^2\theta_{23} = \frac{|U_{\mu3}|^2}{1-|U_{e3}|^2}\,.
  \label{eq:angle_extraction}
\end{equation}
The central values and $1\sigma$ ranges for the solar, atmospheric, and reactor mixing angles  are summarized in Table~\ref{tab:observablesNuFit} from Ref.~\cite{Esteban:2024eli}.
Moreover, by exploiting $U_\mathrm{PMNS}=U_l^\dagger\, U_\nu$ and  setting $U_l=R_{23}(\thetal_{23})$, we obtain the relation:
\begin{equation}
U_\nu = U_l U_{\text{PMNS}}=R_{23}(\theta_{23}^\nu\equiv\theta_{23} + \thetal_{23}) \, R_{13}(\theta_{13}^\nu\equiv\theta_{13}, \delta_\mathrm{CP}) \, R_{12}(\theta_{12}^\nu\equiv\theta_{12})\mathcal{P}.
\label{eq:unu_isolated}
\end{equation}
Consequently, the entire geometric shift induced by $\thetal_{23}$ is absorbed linearly into a single effective neutrino atmospheric angle, $\theta_{23}^\nu$. The remaining mixing angles, namely the solar $\theta_{12}^\nu$ and reactor $\theta_{13}^\nu$ angles, depend exclusively on the parameters of the physical $U_{\text{PMNS}}$ matrix and remain completely untouched by the charged lepton sector rotation.

%%%%%%%%%%%%%%%%%%%%%%%%%%%%%
\section{Neutrino mass matrix and its parametrization}
\label{app:map}
%%%%%%%%%%%%%%%%%%%%%%%%%%%%%

In Eq.~\eqref{eq:FmnuR23}, the neutrino mass matrix has been parametrized in terms of six real parameters $(x, y, z, u, w, q)$. Here, we write explicitly the relations between these effective parameters, the Yukawa couplings, and the Majorana mass terms in Eq.~\ref{eq:diracMajoranaNeutrinos}:
$$\begin{aligned} m_0 &= -\frac{v^2 \varepsilon^2}{\Lambda} \\ q   &= \frac{y_{11}^2 k_{22} - 2 y_{11} y_{12} k_{12} + y_{12}^2 k_{11}}{k_{11} k_{22} - k_{12}^2} \\ u   &= \frac{y_{11} y_{21} k_{22} - (y_{11} y_{22} + y_{12} y_{21}) k_{12} + y_{12} y_{22} k_{11}}{k_{11} k_{22} - k_{12}^2} \\ w   &= \frac{y_{11} y_{31} k_{22} - (y_{11} y_{32} + y_{12} y_{31}) k_{12} + y_{12} y_{32} k_{11}}{k_{11} k_{22} - k_{12}^2} \\ x   &= \frac{y_{21}^2 k_{22} - 2 y_{21} y_{22} k_{12} + y_{22}^2 k_{11}}{k_{11} k_{22} - k_{12}^2} \\ z   &= \frac{y_{21} y_{31} k_{22} - (y_{21} y_{32} + y_{22} y_{31}) k_{12} + y_{22} y_{32} k_{11}}{k_{11} k_{22} - k_{12}^2} \\ y   &= \frac{y_{31}^2 k_{22} - 2 y_{31} y_{32} k_{12} + y_{32}^2 k_{11}}{k_{11} k_{22} - k_{12}^2} \end{aligned}$$

\newpage
\bibliographystyle{JHEP}
\bibliography{refs2}

@article{Batell:2011tc,
    author = "Batell, Brian and Pradler, Josef and Spannowsky, Michael",
    title = "{Dark Matter from Minimal Flavor Violation}",
    eprint = "1105.1781",
    archivePrefix = "arXiv",
    primaryClass = "hep-ph",
    doi = "10.1007/JHEP08(2011)038",
    journal = "JHEP",
    volume = "08",
    pages = "038",
    year = "2011"
}

@article{Calibbi:2020jvd,
    author = "Calibbi, Lorenzo and Redigolo, Diego and Ziegler, Robert and Zupan, Jure",
    title = "{Looking forward to lepton-flavor-violating ALPs}",
    eprint = "2006.04795",
    archivePrefix = "arXiv",
    primaryClass = "hep-ph",
    reportNumber = "P3H-20-024, TTP20-025",
    doi = "10.1007/JHEP09(2021)173",
    journal = "JHEP",
    volume = "09",
    pages = "173",
    year = "2021"
}

@article{Giarnetti:2025idu,
    author = "Giarnetti, A. and Marciano, S. and Meloni, D. and Rettaroli, M.",
    title = "{A Roadmap for neutrino charge assignments in $U(2)_F$ Flavor Models: Implications for LFV processes and leptonic anomalous magnetic moments}",
    eprint = "2505.20281",
    archivePrefix = "arXiv",
    primaryClass = "hep-ph",
    month = "5",
    year = "2025"
}

@article{Ma:2001mr,
    author = "Ma, Ernest and Raidal, Martti",
    title = "{Neutrino mass, muon anomalous magnetic moment, and lepton flavor nonconservation}",
    eprint = "hep-ph/0102255",
    archivePrefix = "arXiv",
    reportNumber = "UCRHEP-T301",
    doi = "10.1103/PhysRevLett.87.011802",
    journal = "Phys. Rev. Lett.",
    volume = "87",
    pages = "011802",
    year = "2001",
    note = "[Erratum: Phys.Rev.Lett. 87, 159901 (2001)]"
}

@article{DUNE:2020ypp,
    author = "Abi, Babak and others",
    collaboration = "DUNE",
    title = "{Deep Underground Neutrino Experiment (DUNE), Far Detector Technical Design Report, Volume II: DUNE Physics}",
    eprint = "2002.03005",
    archivePrefix = "arXiv",
    primaryClass = "hep-ex",
    reportNumber = "FERMILAB-PUB-20-025-ND",
    doi = "10.1088/1748-0221/15/08/T08009",
    journal = "JINST",
    volume = "15",
    number = "08",
    pages = "T08009",
    year = "2020"
}

@article{DUNE:2020jqi,
    author = "Abi, Babak and others",
    collaboration = "DUNE",
    title = "{Long-baseline physics program of the Deep Underground Neutrino Experiment}",
    eprint = "2006.16043",
    archivePrefix = "arXiv",
    primaryClass = "hep-ex",
    reportNumber = "FERMILAB-PUB-20-234-ND",
    doi = "10.1140/epjc/s10052-020-08456-z",
    journal = "Eur. Phys. J. C",
    volume = "80",
    number = "10",
    pages = "978",
    year = "2020"
}

@article{Hyper-Kamiokande:2018ofw,
    author = "Abe, K. and others",
    collaboration = "Hyper-Kamiokande",
    title = "{Hyper-Kamiokande Design Report}",
    eprint = "1805.04163",
    archivePrefix = "arXiv",
    primaryClass = "physics.ins-det",
    reportNumber = "KEK-PREPRINT-2018-2",
    month = "5",
    year = "2018"
}

@article{Schechter:1980gr,
    author = "Schechter, J. and Valle, J. W. F.",
    title = "{Neutrino Masses in SU(2) x U(1) Theories}",
    doi = "10.1103/PhysRevD.22.2227",
    journal = "Phys. Rev. D",
    volume = "22",
    pages = "2227",
    year = "1980"
}

@article{Harrison:1999cf,
    author = "Harrison, P. F. and Perkins, D. H. and Scott, W. G.",
    title = "{A Redetermination of the neutrino mixing matrix establishing the large, nonmaximal u(e2) entry}",
    eprint = "hep-ph/9904297",
    archivePrefix = "arXiv",
    doi = "10.1016/S0370-2693(99)00412-2",
    journal = "Phys. Lett. B",
    volume = "458",
    pages = "79--92",
    year = "1999"
}

@article{Harrison:2002er,
    author = "Harrison, P. F. and Perkins, D. H. and Scott, W. G.",
    title = "{Tri-bimaximal mixing and the neutrino oscillation data}",
    eprint = "hep-ph/0202074",
    archivePrefix = "arXiv",
    doi = "10.1016/S0370-2693(02)01336-9",
    journal = "Phys. Lett. B",
    volume = "530",
    pages = "167",
    year = "2002"
}

@article{Xing:2002sw,
    author = "Xing, Zhi-zhong",
    title = "{Nearly tri-bimaximal neutrino mixing and CP violation}",
    eprint = "hep-ph/0204049",
    archivePrefix = "arXiv",
    doi = "10.1016/S0370-2693(02)01734-3",
    journal = "Phys. Lett. B",
    volume = "533",
    pages = "85--94",
    year = "2002"
}

@article{Vissani:1997pa,
    author = "Vissani, Francesco",
    title = "{A Effective Lagrangian for neutrino masses}",
    eprint = "hep-ph/9709409",
    archivePrefix = "arXiv",
    month = "9",
    year = "1997"
}

@article{Altarelli:2005yx,
    author = "Altarelli, Guido and Feruglio, Ferruccio",
    title = "{Tri-bimaximal neutrino mixing, A(4) gauge symmetry and fermion masses}",
    eprint = "hep-ph/0512103",
    archivePrefix = "arXiv",
    doi = "10.1016/j.nuclphysb.2006.02.015",
    journal = "Nucl. Phys. B",
    volume = "741",
    pages = "215--235",
    year = "2006"
}

@article{King:2011ab,
    author = "King, Stephen F. and Luhn, Christoph",
    title = "{A4 models of tri-bimaximal-reactor neutrino mixing}",
    eprint = "1107.5332",
    archivePrefix = "arXiv",
    primaryClass = "hep-ph",
    doi = "10.1007/JHEP09(2011)042",
    journal = "JHEP",
    volume = "09",
    pages = "042",
    year = "2011"
}

@article{Barger:2001yr,
    author = "Barger, V. and Marfatia, D. and Whisnant, K.",
    title = "{Bimaximal and tri-bimaximal neutrino mixing}",
    eprint = "hep-ph/0104110",
    archivePrefix = "arXiv",
    doi = "10.1016/S0370-2693(01)00424-5",
    journal = "Phys. Lett. B",
    volume = "509",
    pages = "19--29",
    year = "2001"
}

@article{Giarnetti:2024y2407,
    author = "Giarnetti, Alessio and Marciano, Simone and Meloni, Davide",
    title = "{Exploring new physics with DUNE high energy flux: the case of Lorentz Invariance Violation, Large Extra Dimensions and Long Range Forces}",
    eprint = "2407.17247",
    archivePrefix = "arXiv",
    primaryClass = "hep-ph",
    journal = "Universe",
    volume = "10",
    number = "9",
    pages = "357",
    year = "2024"
}

@article{ESSnuSB:2025y2504,
    author = "Aguilar, J. and others",
    collaboration = "ESSnuSB",
    title = "{Probing Long-Range Forces in Neutrino Oscillations at the ESSnuSB Experiment}",
    eprint = "2504.10480",
    archivePrefix = "arXiv",
    primaryClass = "hep-ph",
    journal = "JHEP",
    volume = "07",
    pages = "186",
    year = "2025"
}

@article{Giarnetti:2026nbd,
    author = "Giarnetti, Alessio and Marciano, Simone and Meloni, Davide",
    title = "{Constraining the Neutrino Mixing Matrix via Single-Sector Charged-Lepton Rotations in the JUNO Precision Era}",
    eprint = "2603.29876",
    archivePrefix = "arXiv",
    primaryClass = "hep-ph",
    doi = "10.3390/sym18060954",
    journal = "Symmetry",
    volume = "18",
    number = "6",
    pages = "954",
    year = "2026"
}

@inbook{Feruglio:2017spp,
    author = "Feruglio, Ferruccio",
    editor = "Levy, Aharon and Forte, Stefano and Ridolfi, Giovanni",
    title = "{Are neutrino masses modular forms?}",
    booktitle = "{From My Vast Repertoire ...}: {Guido Altarelli's Legacy}",
    eprint = "1706.08749",
    archivePrefix = "arXiv",
    primaryClass = "hep-ph",
    reportNumber = "DFPD-2017-TH-09",
    doi = "10.1142/9789813238053_0012",
    pages = "227--266",
    year = "2019"
}

@article{Beniwal:2020hjc,
    author = "Beniwal, Ankit and Herrero-Garc{\'\i}a, Juan and Leerdam, Nicholas and White, Martin and Williams, Anthony G.",
    title = "{The ScotoSinglet Model: a scalar singlet extension of the Scotogenic Model}",
    eprint = "2010.05937",
    archivePrefix = "arXiv",
    primaryClass = "hep-ph",
    reportNumber = "ADP-20-25/T1135, CP3-20-46, IFIC/20-46",
    doi = "10.1007/JHEP06(2021)136",
    journal = "JHEP",
    volume = "21",
    pages = "136",
    year = "2020"
}

@article{King:2013iva,
    author = "King, Stephen F.",
    title = "{Minimal predictive see-saw model with normal neutrino mass hierarchy}",
    eprint = "1304.6264",
    archivePrefix = "arXiv",
    primaryClass = "hep-ph",
    doi = "10.1007/JHEP07(2013)137",
    journal = "JHEP",
    volume = "07",
    pages = "137",
    year = "2013"
}

@article{Marciano:2024nwm,
    author = "Marciano, Simone and Meloni, Davide and Parriciatu, Matteo",
    title = "{Minimal seesaw and leptogenesis with the smallest modular finite group}",
    eprint = "2402.18547",
    archivePrefix = "arXiv",
    primaryClass = "hep-ph",
    doi = "10.1007/JHEP05(2024)020",
    journal = "JHEP",
    volume = "05",
    pages = "020",
    year = "2024"
}

@article{Xing:2020ald,
    author = "Xing, Zhi-zhong and Zhao, Zhen-hua",
    title = "{The minimal seesaw and leptogenesis models}",
    eprint = "2008.12090",
    archivePrefix = "arXiv",
    primaryClass = "hep-ph",
    doi = "10.1088/1361-6633/abf086",
    journal = "Rept. Prog. Phys.",
    volume = "84",
    number = "6",
    pages = "066201",
    year = "2021"
}

@article{Wang:2019ovr,
    author = "Wang, Xin and Zhou, Shun",
    title = "{The minimal seesaw model with a modular S$_{4}$ symmetry}",
    eprint = "1910.09473",
    archivePrefix = "arXiv",
    primaryClass = "hep-ph",
    doi = "10.1007/JHEP05(2020)017",
    journal = "JHEP",
    volume = "05",
    pages = "017",
    year = "2020"
}

@article{Escribano:2020iqq,
    author = "Escribano, Pablo and Reig, Mario and Vicente, Avelino",
    title = "{Generalizing the Scotogenic model}",
    eprint = "2004.05172",
    archivePrefix = "arXiv",
    primaryClass = "hep-ph",
    reportNumber = "IFIC/20-13",
    doi = "10.1007/JHEP07(2020)097",
    journal = "JHEP",
    volume = "07",
    pages = "097",
    year = "2020"
}

@article{Hagedorn:2018spx,
    author = "Hagedorn, Claudia and Herrero-Garc{\'\i}a, Juan and Molinaro, Emiliano and Schmidt, Michael A.",
    title = "{Phenomenology of the Generalised Scotogenic Model with Fermionic Dark Matter}",
    eprint = "1804.04117",
    archivePrefix = "arXiv",
    primaryClass = "hep-ph",
    reportNumber = "ADP-18-9-T1057, CP3-ORIGINS-2018-012, ADP-18-9/T1057, CP3-Origins-2018-012-DNRF90",
    doi = "10.1007/JHEP11(2018)103",
    journal = "JHEP",
    volume = "11",
    pages = "103",
    year = "2018"
}

@article{Ma:2006km,
    author = "Ma, Ernest",
    title = "{Verifiable radiative seesaw mechanism of neutrino mass and dark matter}",
    eprint = "hep-ph/0601225",
    archivePrefix = "arXiv",
    reportNumber = "UCRHEP-T403",
    doi = "10.1103/PhysRevD.73.077301",
    journal = "Phys. Rev. D",
    volume = "73",
    pages = "077301",
    year = "2006"
}

@article{Giarnetti:2023osf,
    author = "Giarnetti, Alessio and Herrero-Garc{\'\i}a, Juan and Marciano, Simone and Meloni, Davide and Vatsyayan, Drona",
    title = "{Neutrino masses from new seesaw models: low-scale variants and phenomenological implications}",
    eprint = "2312.14119",
    archivePrefix = "arXiv",
    primaryClass = "hep-ph",
    doi = "10.1140/epjc/s10052-024-13149-y",
    journal = "Eur. Phys. J. C",
    volume = "84",
    number = "8",
    pages = "803",
    year = "2024"
}

@article{Cai:2017jrq,
    author = "Cai, Yi and Herrero-Garc{\'\i}a, Juan and Schmidt, Michael A. and Vicente, Avelino and Volkas, Raymond R.",
    title = "{From the trees to the forest: a review of radiative neutrino mass models}",
    eprint = "1706.08524",
    archivePrefix = "arXiv",
    primaryClass = "hep-ph",
    reportNumber = "ADP-17-29-T1035",
    doi = "10.3389/fphy.2017.00063",
    journal = "Front. in Phys.",
    volume = "5",
    pages = "63",
    year = "2017"
}

@article{Restrepo:2013aga,
    author = "Restrepo, Diego and Zapata, Oscar and Yaguna, Carlos E.",
    title = "{Models with radiative neutrino masses and viable dark matter candidates}",
    eprint = "1308.3655",
    archivePrefix = "arXiv",
    primaryClass = "hep-ph",
    doi = "10.1007/JHEP11(2013)011",
    journal = "JHEP",
    volume = "11",
    pages = "011",
    year = "2013"
}

@article{Arbelaez:2022ejo,
    author = "Arbel{\'a}ez, Carolina and Cepedello, Ricardo and Helo, Juan Carlos and Hirsch, Martin and Kovalenko, Sergey",
    title = "{How many 1-loop neutrino mass models are there?}",
    eprint = "2205.13063",
    archivePrefix = "arXiv",
    primaryClass = "hep-ph",
    doi = "10.1007/JHEP08(2022)023",
    journal = "JHEP",
    volume = "08",
    pages = "023",
    year = "2022"
}

@article{JUNO:2015zny,
    author = "An, Fengpeng and others",
    collaboration = "JUNO",
    title = "{Neutrino Physics with JUNO}",
    eprint = "1507.05613",
    archivePrefix = "arXiv",
    primaryClass = "physics.ins-det",
    doi = "10.1088/0954-3899/43/3/030401",
    journal = "J. Phys. G",
    volume = "43",
    number = "3",
    pages = "030401",
    year = "2016"
}

@article{Super-Kamiokande:2017yvm,
    author = "Abe, K. and others",
    collaboration = "Super-Kamiokande",
    title = "{Atmospheric neutrino oscillation analysis with external constraints in Super-Kamiokande I-IV}",
    eprint = "1710.09126",
    archivePrefix = "arXiv",
    primaryClass = "hep-ex",
    doi = "10.1103/PhysRevD.97.072001",
    journal = "Phys. Rev. D",
    volume = "97",
    number = "7",
    pages = "072001",
    year = "2018"
}

@article{Antusch:2003kp,
    author = {Antusch, Stefan and Kersten, J{\"o}rn and Lindner, Manfred and Ratz, Michael},
    title = "{Running neutrino masses, mixings and CP phases: Analytical results and phenomenological consequences}",
    eprint = "hep-ph/0305273",
    archivePrefix = "arXiv",
    reportNumber = "TUM-HEP-510-03, DESY-03-065",
    doi = "10.1016/j.nuclphysb.2003.09.050",
    journal = "Nucl. Phys. B",
    volume = "674",
    pages = "401--433",
    year = "2003"
}

@article{Mohapatra:2005ra,
    author = "Mohapatra, R. N. and Nasri, S. and Yu, Hai-Bo",
    title = "{Leptogenesis, $\mu$ - $\tau$ symmetry and $\theta_{13}$}",
    eprint = "hep-ph/0502026",
    archivePrefix = "arXiv",
    reportNumber = "UMD-PP-05-040",
    doi = "10.1016/j.physletb.2005.03.082",
    journal = "Phys. Lett. B",
    volume = "615",
    pages = "231--239",
    year = "2005"
}

@article{Arcadi:2022ojj,
    author = "Arcadi, Giorgio and Marciano, Simone and Meloni, Davide",
    title = "{Neutrino mixing and leptogenesis in a $L_e-L_\mu -L_\tau $ model}",
    eprint = "2205.02565",
    archivePrefix = "arXiv",
    primaryClass = "hep-ph",
    doi = "10.1140/epjc/s10052-023-11268-6",
    journal = "Eur. Phys. J. C",
    volume = "83",
    number = "2",
    pages = "137",
    year = "2023"
}

@article{Falkowski:2015zwa,
    author = "Falkowski, Adam and Nardecchia, Marco and Ziegler, Robert",
    title = "{Lepton Flavor Non-Universality in B-meson Decays from a U(2) Flavor Model}",
    eprint = "1509.01249",
    archivePrefix = "arXiv",
    primaryClass = "hep-ph",
    doi = "10.1007/JHEP11(2015)173",
    journal = "JHEP",
    volume = "11",
    pages = "173",
    year = "2015"
}

@article{Barbieri:1997tu,
    author = "Barbieri, Riccardo and Hall, Lawrence J. and Romanino, Andrea",
    title = "{Consequences of a U(2) flavor symmetry}",
    eprint = "hep-ph/9702315",
    archivePrefix = "arXiv",
    reportNumber = "IFUP-TH-4-97, LBL-39946, LBNL-39946, UCB-PTH-97-06",
    doi = "10.1016/S0370-2693(97)00372-9",
    journal = "Phys. Lett. B",
    volume = "401",
    pages = "47--53",
    year = "1997"
}

@article{Barbieri:1995uv,
    author = "Barbieri, Riccardo and Dvali, G. R. and Hall, Lawrence J.",
    title = "{Predictions from a U(2) flavor symmetry in supersymmetric theories}",
    eprint = "hep-ph/9512388",
    archivePrefix = "arXiv",
    reportNumber = "LBL-38065, UCB-PTH-95-44",
    doi = "10.1016/0370-2693(96)00318-8",
    journal = "Phys. Lett. B",
    volume = "377",
    pages = "76--82",
    year = "1996"
}

@article{Esteban:2024eli,
    author = "Esteban, Ivan and Gonzalez-Garcia, M. C. and Maltoni, Michele and Martinez-Soler, Ivan and Pinheiro, Jo{\~a}o Paulo and Schwetz, Thomas",
    title = "{NuFit-6.0: updated global analysis of three-flavor neutrino oscillations}",
    eprint = "2410.05380",
    archivePrefix = "arXiv",
    primaryClass = "hep-ph",
    reportNumber = "IFT-UAM/CSIC-24-140, YITP-SB-2024-24, IPPP/24/64, IPPP/24/64, IFT-UAM/CSIC-24-140, YITP-SB-2024-24",
    doi = "10.1007/JHEP12(2024)216",
    journal = "JHEP",
    volume = "12",
    pages = "216",
    year = "2024"
}

@article{Calibbi:2025rxn,
    author = "Calibbi, Lorenzo and Yi, Jiangyi",
    title = "{Phenomenology of Non-Abelian Gauge and Goldstone Bosons in a U(2) Flavor Model}",
    eprint = "2511.10468",
    archivePrefix = "arXiv",
    primaryClass = "hep-ph",
    month = "11",
    year = "2025"
}

@article{Linster:2018avp,
    author = "Linster, Matthias and Ziegler, Robert",
    title = "{A Realistic $U(2)$ Model of Flavor}",
    eprint = "1805.07341",
    archivePrefix = "arXiv",
    primaryClass = "hep-ph",
    reportNumber = "CERN-TH-2018-121, TTP18-019",
    doi = "10.1007/JHEP08(2018)058",
    journal = "JHEP",
    volume = "08",
    pages = "058",
    year = "2018"
}

@article{Mandal:2023jnv,
    author = "Mandal, Rusa and Tong, Tom",
    title = "{Exploring freeze-out and freeze-in dark matter via effective Froggatt-Nielsen theory}",
    eprint = "2307.14972",
    archivePrefix = "arXiv",
    primaryClass = "hep-ph",
    doi = "10.1088/1475-7516/2023/11/074",
    journal = "JCAP",
    volume = "11",
    pages = "074",
    year = "2023"
}

@article{ParticleData:2026uvj,
  author = {Takahashi, F. and others},
  collaboration = {Particle Data},
  title = {{Review of Particle Physics*}},
  doi = {10.1142/s0217751x26300115},
  journal = {Int. J. Mod. Phys. A},
  volume = {41},
  number = {22},
  pages = {2630011},
  year = {2026},
}

@article{Feruglio:2015jfa,
  author = {Feruglio, Ferruccio},
  title = {{Pieces of the Flavour Puzzle}},
  eprint = {1503.04071},
  archiveprefix = {arXiv},
  primaryclass = {hep-ph},
  doi = {10.1140/epjc/s10052-015-3576-5},
  journal = {Eur. Phys. J. C},
  volume = {75},
  number = {8},
  pages = {373},
  year = {2015},
}

@article{Barbieri:2011ci,
  author = {Barbieri, Riccardo and Isidori, Gino and Jones-Perez, Joel and Lodone, Paolo and Straub, David M.},
  title = {{$U(2)$ and Minimal Flavour Violation in Supersymmetry}},
  eprint = {1105.2296},
  archiveprefix = {arXiv},
  primaryclass = {hep-ph},
  doi = {10.1140/epjc/s10052-011-1725-z},
  journal = {Eur. Phys. J. C},
  volume = {71},
  pages = {1725},
  year = {2011},
}

@article{Isidori:2012ts,
  author = {Isidori, Gino and Straub, David M.},
  title = {{Minimal Flavour Violation and Beyond}},
  eprint = {1202.0464},
  archiveprefix = {arXiv},
  primaryclass = {hep-ph},
  doi = {10.1140/epjc/s10052-012-2103-1},
  journal = {Eur. Phys. J. C},
  volume = {72},
  pages = {2103},
  year = {2012},
}

@article{Lopez-Honorez:2013wla,
  author = {Lopez-Honorez, Laura and Merlo, Luca},
  title = {{Dark matter within the minimal flavour violation ansatz}},
  eprint = {1303.1087},
  archiveprefix = {arXiv},
  primaryclass = {hep-ph},
  reportnumber = {FTUAM-13-129, IFT-UAM-CSIC-13-018, CERN-PH-TH-2013-034},
  doi = {10.1016/j.physletb.2013.04.015},
  journal = {Phys. Lett. B},
  volume = {722},
  pages = {135--143},
  year = {2013},
}

@article{Mescia:2024rki,
  author = {Mescia, Federico and Okawa, Shohei and Wu, Keyun},
  title = {{Multi-component dark matter from Minimal Flavor Violation}},
  eprint = {2408.16812},
  archiveprefix = {arXiv},
  primaryclass = {hep-ph},
  reportnumber = {KEK-TH-2649},
  doi = {10.1007/JHEP11(2024)114},
  journal = {JHEP},
  volume = {11},
  pages = {114},
  year = {2024},
}

@article{Mescia:2026xju,
  author = {Mescia, Federico and Okawa, Shohei and Swallow, Joel and Toni, Claudio},
  title = {{Dark matter emission at Belle II and NA62 in the minimal flavor violation framework}},
  eprint = {2601.08907},
  archiveprefix = {arXiv},
  primaryclass = {hep-ph},
  reportnumber = {KYUSHU-HET-349, LAPTH-004/26},
  doi = {10.1103/7dlz-bwlz},
  journal = {Phys. Rev. D},
  volume = {113},
  number = {9},
  pages = {095014},
  year = {2026},
}

@article{Cirelli:2024ssz,
  author = {Cirelli, Marco and Strumia, Alessandro and Zupan, Jure},
  title = {{Dark Matter}},
  eprint = {2406.01705},
  archiveprefix = {arXiv},
  primaryclass = {hep-ph},
  doi = {10.21468/SciPostPhysRev.1},
  month = {6},
  year = {2024},
}

@article{Bertone:2016nfn,
  author = {Bertone, Gianfranco and Hooper, Dan},
  title = {{History of dark matter}},
  eprint = {1605.04909},
  archiveprefix = {arXiv},
  primaryclass = {astro-ph.CO},
  reportnumber = {FERMILAB-PUB-16-157-A},
  doi = {10.1103/RevModPhys.90.045002},
  journal = {Rev. Mod. Phys.},
  volume = {90},
  number = {4},
  pages = {045002},
  year = {2018},
}

@article{Bertone:2004pz,
  author = {Bertone, Gianfranco and Hooper, Dan and Silk, Joseph},
  title = {{Particle dark matter: Evidence, candidates and constraints}},
  eprint = {hep-ph/0404175},
  archiveprefix = {arXiv},
  reportnumber = {FERMILAB-PUB-04-047-A},
  doi = {10.1016/j.physrep.2004.08.031},
  journal = {Phys. Rept.},
  volume = {405},
  pages = {279--390},
  year = {2005},
}

@article{Peter:2012rz,
  author = {Peter, Annika H. G.},
  title = {{Dark Matter: A Brief Review}},
  eprint = {1201.3942},
  archiveprefix = {arXiv},
  primaryclass = {astro-ph.CO},
  month = {1},
  year = {2012},
}

@article{Leurer:1993gy,
    author = "Leurer, Miriam and Nir, Yosef and Seiberg, Nathan",
    title = "{Mass matrix models: The Sequel}",
    eprint = "hep-ph/9310320",
    archivePrefix = "arXiv",
    reportNumber = "RU-93-43, WIS-93-93-PH",
    doi = "10.1016/0550-3213(94)90074-4",
    journal = "Nucl. Phys. B",
    volume = "420",
    pages = "468--504",
    year = "1994"
}

@article{Calibbi:2015sfa,
    author = "Calibbi, Lorenzo and Crivellin, Andreas and Zald{\'\i}var, Bryan",
    title = "{Flavor portal to dark matter}",
    eprint = "1501.07268",
    archivePrefix = "arXiv",
    primaryClass = "hep-ph",
    reportNumber = "ULB-TH-15-01, CERN-PH-TH-2015-010",
    doi = "10.1103/PhysRevD.92.016004",
    journal = "Phys. Rev. D",
    volume = "92",
    number = "1",
    pages = "016004",
    year = "2015"
}

@article{Alvarado:2017bax,
    author = "Alvarado, Carlos and Elahi, Fatemeh and Raj, Nirmal",
    title = "{Thermal dark matter via the flavon portal}",
    eprint = "1706.03081",
    archivePrefix = "arXiv",
    primaryClass = "hep-ph",
    doi = "10.1103/PhysRevD.96.075002",
    journal = "Phys. Rev. D",
    volume = "96",
    number = "7",
    pages = "075002",
    year = "2017"
}

@article{Cheek:2022yof,
    author = "Cheek, Andrew and Osi{\'n}ski, Jacek K. and Roszkowski, Leszek and Trojanowski, Sebastian",
    title = "{Dark matter production through a non-thermal flavon portal}",
    eprint = "2211.02057",
    archivePrefix = "arXiv",
    primaryClass = "hep-ph",
    doi = "10.1007/JHEP03(2023)149",
    journal = "JHEP",
    volume = "03",
    pages = "149",
    year = "2023"
}

@article{Babu:2023zni,
    author = "Babu, K. S. and Chakdar, Shreyashi and Das, Nandini and Ghosh, Dilip Kumar and Ghosh, Purusottam",
    title = "{FIMP dark matter from flavon portals}",
    eprint = "2305.03167",
    archivePrefix = "arXiv",
    primaryClass = "hep-ph",
    doi = "10.1007/JHEP07(2023)143",
    journal = "JHEP",
    volume = "07",
    pages = "143",
    year = "2023"
}

@article{BasBeneito2022,
  author = {Bas i Beneito, Arnau and Herrero-García, Juan and Vatsyayan, Drona},
  title = {Multi-component dark sectors: symmetries, asymmetries and conversions},
  journal = {JHEP},
  volume = {10},
  pages = {075},
  year = {2022},
  doi = {10.1007/JHEP10(2022)075},
  eprint = {2207.02874},
  archivePrefix = {arXiv}
}

@article{YagunaZapata2020,
  author = {Yaguna, Carlos E. and Zapata, Oscar},
  title = {Multi-component scalar dark matter from a $Z_N$ symmetry: a systematic analysis},
  journal = {JHEP},
  volume = {03},
  pages = {109},
  year = {2020},
  doi = {10.1007/JHEP03(2020)109},
  eprint = {1911.05515},
  archivePrefix = {arXiv}
}

@article{BelangerMjallalPukhov2022,
  author = {Bélanger, Geneviève and Mjallal, A. and Pukhov, A.},
  title = {Two dark matter candidates: the case of inert doublet and singlet scalars},
  journal = {Phys. Rev. D},
  volume = {105},
  pages = {035018},
  year = {2022},
  doi = {10.1103/PhysRevD.105.035018},
  eprint = {2108.08061},
  archivePrefix = {arXiv}
}

@article{Martin1992,
  author = {Martin, Stephen P.},
  title = {Some simple criteria for gauged {R}-parity},
  journal = {Phys. Rev. D},
  volume = {46},
  pages = {R2769},
  year = {1992},
  doi = {10.1103/PhysRevD.46.R2769},
  eprint = {hep-ph/9207218},
  archivePrefix = {arXiv}
}

@article{Martin1996,
  author = {Martin, Stephen P.},
  title = {Implications of supersymmetric models with natural {R}-parity conservation},
  journal = {Phys. Rev. D},
  volume = {54},
  pages = {2340--2348},
  year = {1996},
  doi = {10.1103/PhysRevD.54.2340},
  eprint = {hep-ph/9602349},
  archivePrefix = {arXiv}
}

@article{HerreroGarciaLandiniYanagida2025,
  author = {Herrero-García, Juan and Landini, Giacomo and Yanagida, Tsutomu T.},
  title = {Dark matter in the high-scale seesaw leptogenesis paradigm},
  journal = {Phys. Rev. D},
  volume = {111},
  pages = {075033},
  year = {2025},
  doi = {10.1103/PhysRevD.111.075033},
  eprint = {2411.03452},
  archivePrefix = {arXiv}
}

@article{Elahi:2014fsa,
    author = "Elahi, Fatemeh and Kolda, Christopher and Unwin, James",
    title = "{UltraViolet Freeze-in}",
    eprint = "1410.6157",
    archivePrefix = "arXiv",
    primaryClass = "hep-ph",
    doi = "10.1007/JHEP03(2015)048",
    journal = "JHEP",
    volume = "03",
    pages = "048",
    year = "2015"
}

@article{Froggatt:1978nt,
    author = "Froggatt, C. D. and Nielsen, Holger Bech",
    title = "{Hierarchy of Quark Masses, Cabibbo Angles and CP Violation}",
    reportNumber = "CERN-TH-2519",
    doi = "10.1016/0550-3213(79)90316-X",
    journal = "Nucl. Phys. B",
    volume = "147",
    pages = "277--298",
    year = "1979"
}

@article{Leurer:1992wg,
    author = "Leurer, Miriam and Nir, Yosef and Seiberg, Nathan",
    title = "{Mass matrix models}",
    eprint = "hep-ph/9212278",
    archivePrefix = "arXiv",
    reportNumber = "RU-92-59, WIS-92-94-PH",
    doi = "10.1016/0550-3213(93)90112-3",
    journal = "Nucl. Phys. B",
    volume = "398",
    pages = "319--342",
    year = "1993"
}

@article{Marzocca2013b,
  title = {Nonzero $|U_{e3} |$ from charged lepton corrections and the atmospheric neutrino mixing angle},
  volume = {2013},
  issn = {1029-8479},
  url = {http://dx.doi.org/10.1007/JHEP05(2013)073},
  doi = {10.1007/jhep05(2013)073},
  number = {5},
  journal = {Journal of High Energy Physics},
  publisher = {Springer Science and Business Media LLC},
  author = {Marzocca, David and Petcov, S. T. and Romanino, Andrea and Sevilla, M. C.},
  year = {2013},
  month = May,
}

@article{AristizabalSierra:2014irc,
  author = {Aristizabal Sierra, D. and Dhen, Mika{\"e}l and Fong, Chee Sheng and Vicente, Avelino},
  title = {{Dynamical flavor origin of $\mathbb{Z}_N$ symmetries}},
  eprint = {1412.5600},
  archiveprefix = {arXiv},
  primaryclass = {hep-ph},
  doi = {10.1103/PhysRevD.91.096004},
  journal = {Phys. Rev. D},
  volume = {91},
  number = {9},
  pages = {096004},
  year = {2015},
}

@article{Planck:2018vyg,
  author = {Aghanim, N. and others},
  collaboration = {Planck},
  title = {{Planck 2018 results. VI. Cosmological parameters}},
  eprint = {1807.06209},
  archiveprefix = {arXiv},
  primaryclass = {astro-ph.CO},
  doi = {10.1051/0004-6361/201833910},
  journal = {Astron. Astrophys.},
  volume = {641},
  pages = {A6},
  year = {2020},
  note = {[Erratum: Astron.Astrophys. 652, C4 (2021)]},
}

@article{Buttazzo:2019mvl,
    author = "Buttazzo, Dario and Di Luzio, Luca and Ghorbani, Parsa and Gross, Christian and Landini, Giacomo and Strumia, Alessandro and Teresi, Daniele and Wang, Jin-Wei",
    title = "{Scalar gauge dynamics and Dark Matter}",
    eprint = "1911.04502",
    archivePrefix = "arXiv",
    primaryClass = "hep-ph",
    reportNumber = "DESY 19-197, DESY-19-197",
    doi = "10.1007/JHEP01(2020)130",
    journal = "JHEP",
    volume = "01",
    pages = "130",
    year = "2020"
}

@article{Bertolini:2020hjc,
  author = {Bertolini, Stefano and Di Luzio, Luca and Nesti, Fabrizio},
  title = {{Axion-mediated forces, CP violation and left-right interactions}},
  eprint = {2006.12508},
  archiveprefix = {arXiv},
  primaryclass = {hep-ph},
  reportnumber = {DESY 20-103, DESY-20-103},
  doi = {10.1103/PhysRevLett.126.081801},
  journal = {Phys. Rev. Lett.},
  volume = {126},
  number = {8},
  pages = {081801},
  year = {2021},
}

@article{Hall:2009bx,
    author = "Hall, Lawrence J. and Jedamzik, Karsten and March-Russell, John and West, Stephen M.",
    title = "{Freeze-In Production of FIMP Dark Matter}",
    eprint = "0911.1120",
    archivePrefix = "arXiv",
    primaryClass = "hep-ph",
    reportNumber = "OUTP-09-18-P, UCB-PTH-09-32",
    doi = "10.1007/JHEP03(2010)080",
    journal = "JHEP",
    volume = "03",
    pages = "080",
    year = "2010"
}

@article{Calibbi:2016hwq,
    author = "Calibbi, Lorenzo and Goertz, Florian and Redigolo, Diego and Ziegler, Robert and Zupan, Jure",
    title = "{Minimal axion model from flavor}",
    eprint = "1612.08040",
    archivePrefix = "arXiv",
    primaryClass = "hep-ph",
    reportNumber = "TTP16-058, CERN-TH-2016-261",
    doi = "10.1103/PhysRevD.95.095009",
    journal = "Phys. Rev. D",
    volume = "95",
    number = "9",
    pages = "095009",
    year = "2017"
}

@article{Ema:2016ops,
    author = "Ema, Yohei and Hamaguchi, Koichi and Moroi, Takeo and Nakayama, Kazunori",
    title = "{Flaxion: a minimal extension to solve puzzles in the standard model}",
    eprint = "1612.05492",
    archivePrefix = "arXiv",
    primaryClass = "hep-ph",
    reportNumber = "UT-16-36, IPMU16-0189",
    doi = "10.1007/JHEP01(2017)096",
    journal = "JHEP",
    volume = "01",
    pages = "096",
    year = "2017"
}

@article{DEramo:2023nzt,
    author = "D'Eramo, Francesco and Hajkarim, Fazlollah and Lenoci, Alessandro",
    title = "{Dark radiation from the primordial thermal bath in momentum space}",
    eprint = "2311.04974",
    archivePrefix = "arXiv",
    primaryClass = "hep-ph",
    reportNumber = "DESY-23-177",
    doi = "10.1088/1475-7516/2024/03/009",
    journal = "JCAP",
    volume = "03",
    pages = "009",
    year = "2024"
}
\end{document}